%% file: main-arxiv.tex
\documentclass{article}
\usepackage{graphicx} 
\usepackage{xcolor} 
\usepackage{tikz}
\usetikzlibrary{arrows.meta,positioning,shapes.geometric,fit}
\usepackage{tabularx}
\usepackage{url}

\usepackage[letterpaper, margin=1in]{geometry}

\usepackage{tikz}
\usetikzlibrary{arrows.meta,positioning,calc,fit}

\usepackage[most]{tcolorbox}
\tcbuselibrary{skins}

\title{Quantum-Resistant Networks:\\ 
A Review of Primitives, Protocols and Best Practices}
\date{}

\author{
  Elisa Bertino\\
  Purdue University
  \and
  Ramana Kompella\\
  Cisco Research
  \and
  Ashish Kundu\\
  Cisco Research
  \and Cristina Nita-Rotaru\\
  Northeastern University
  \and Jaideep Vaidya\\
  Rutgers University
  \and
  Attila A. Yavuz\\
  University of South Florida
}

\begin{document}

\maketitle
\begin{abstract}
\input{abstract}
\end{abstract}


\input{introduction}
\begin{figure*}[t]
\centering
\includegraphics[width=\textwidth]{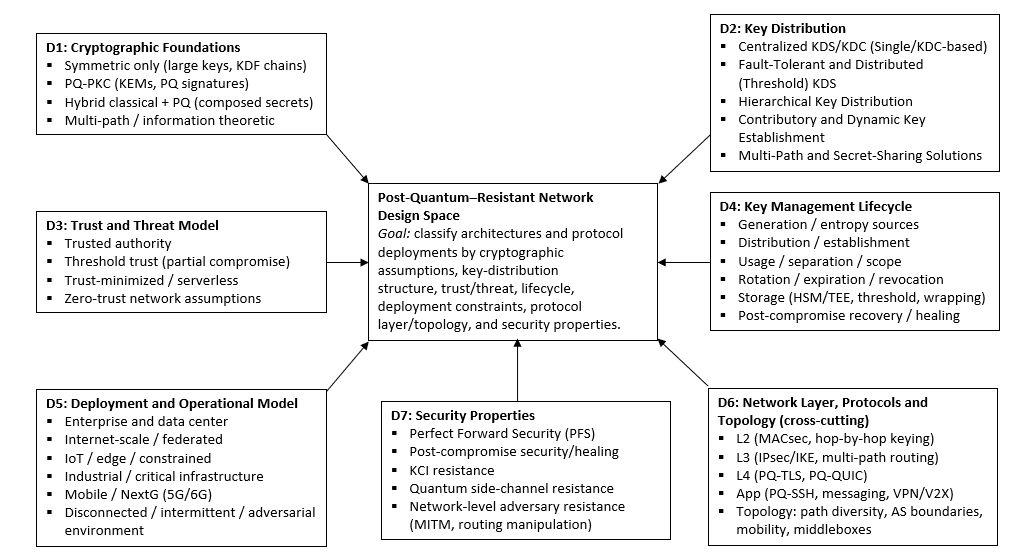}
\caption{``Taxonomy Map'' highlighting the paper’s dimensions and how they jointly structure the PQ network design space.}
\label{taxonomy-map}
\end{figure*}
\input{relatedwork}

\input{CryptographicFoundations}
\input{KeyDistributionArchitectures}
\input{TrustThreatModel}

\input{KeyManagementLifecyle}

\input{DeploymentOperationalModel}

\input{CommunicationTopologyNetworkLayer}
\input{SecurityProperties}
\input{BestPractices}
\section{Concluding Remarks}
\input{ConcludingRemarks}



\bibliographystyle{plain}
\bibliography{References}

\input{Appendix}

\end{document}

%% file: abstract.tex
The advent of large-scale quantum computers threatens the public-key cryptographic foundations that underpin today’s network security infrastructures. While significant progress has been made in standardizing post-quantum cryptographic (PQC) primitives and adapting individual protocols such as TLS and SSH, far less attention has been paid to the broader architectural consequences of the post-quantum transition for networked systems. In particular, many real-world deployments such as mobile networks, industrial control systems, IoT environments, and regulated infrastructures, cannot assume the universal availability, deployability, or desirability of PQ public-key infrastructures.
This paper presents the first comprehensive systematization of PQ–resistant network architectures, focusing on key distribution and management as a system-level design problem rather than a protocol-local substitution. We introduce a unified taxonomy spanning cryptographic foundations (symmetric-only, PQ-PKI, hybrid, and information-theoretic multi-path), key-distribution architectures (centralized, hierarchical, replicated, threshold, MPC-backed, and serverless), trust and threat models, key-management lifecycle, and deployment environments. Using this framework, we analyze the security, scalability, and operational trade-offs of a wide range of architectures under realistic PQ adversary assumptions, including ``harvest-now, decrypt-later'' attacks and partial infrastructure compromise. Our study highlights fundamental gaps in existing approaches, clarifies when PQ-PKI is necessary or avoidable, and identifies promising research directions for building cryptographically agile, quantum-resilient network infrastructures. Building on this systematization, the paper distills a set of best practices for quantum-resistant network design, emphasizing architectural cryptographic agility, lifecycle-aware key management, compromise containment through threshold and distributed trust, and the careful alignment of cryptographic mechanisms with deployment constraints. These practices provide concrete guidance for practitioners navigating heterogeneous and long-lived post-quantum transition environments. 

%% file: introduction.tex
\section{Introduction}

The security of modern communication networks relies fundamentally on cryptographic protocols that enable authentication, key establishment, and secure channel formation. Today’s and tomorrow's infrastructures, from enterprise networks and cloud platforms to industrial control systems, mobile networks, and IoT deployments, all depend heavily on public-key cryptography (PKC), such as RSA, Diffie–Hellman (DH), and elliptic-curve constructions (ECC). These mechanisms underpin protocols including TLS, QUIC, IPsec, SSH, and Kerberos, and form the basis of global trust infrastructures such as PKI and certificate-based authentication. However, the advent of large-scale quantum computers threatens to render classical PKC obsolete. Shor’s algorithm breaks RSA, DH, and ECC entirely, while Grover’s algorithm weakens symmetric constructions unless larger keys are used. As a result, network security architectures must undergo a fundamental transformation to remain secure in the post-quantum (PQ) era.

In response to such very much required transition, considerable progress has been made on post-quantum cryptographic (PQC) primitives, standardization efforts, and protocol-level adaptations. NIST has standardized lattice-based key encapsulation mechanisms (KEMs) and digital signatures, while numerous proposals have explored PQ-TLS, hybrid key exchange, PQ-SSH, and PQ-IPsec. Yet these efforts primarily address ``protocol instantiations'', that is, how to replace or augment individual algorithms inside existing endpoints. Much less attention has been given to the architectural and systems-level consequences of the PQ transition: What happens to networks that rely heavily on symmetric trust anchors? How should centralized or hierarchical key-distribution systems evolve? What if a deployment environment cannot assume the availability of PQ-PKC? And how can multi-path, symmetric-only, or trust-minimized approaches provide quantum-resilient key management without relying on PKI at all? 
Also what security properties can be guaranteed when hybrid schemes are employed, given that their effective security depends on how they are composed and how keys are combined and managed over their lifecycle?

These questions highlight a broader challenge: existing guidelines and SoK efforts focus largely on cryptographic primitives or specific protocols, leaving a significant gap in the understanding of \emph{quantum-safe network architectures} and their security. In many real-world systems—including mobile core networks, industrial deployments, constrained IoT environments, highly regulated sectors, PKI may be unavailable, impractical, too expensive to deploy, or undesirable due to operational assumptions. Such environments require alternative, equally rigorous, quantum-resistant approaches: symmetric-only designs, threshold and multi-party computation (MPC)-backed key-distribution servers (KDSs), replicated or hierarchical KDS architectures, multi-path secret-sharing mechanisms, and hybrid cryptographic strategies that balance resource constraints, performance, deployability, and security.

To our knowledge, no existing survey systematically examines PQ-resistant network and key management architectures across symmetric-only, PQ-PKI, hybrid, distributed, and serverless multi-path models. Prior work does not 
integrate threshold-MPC-backed KDSs, hierarchical trust infrastructures, replicated key distribution center (KDC) clusters, or symmetric-only and multi-path secret-sharing frameworks into a unified taxonomy. Thus, existing surveys leave open the fundamental architectural question: \emph{How should networks distribute and manage cryptographic keys securely in a world where post-quantum public-key cryptography (PQ-PKC) may or may not be available, practical, or trusted?}

This paper fills that gap by providing the first survey that spans the architectural, trust-model, and distribution-mechanism layers of quantum-resistant network design, offering a comprehensive taxonomy (see Fig.~\ref{taxonomy-map}) that connects symmetric primitives, PQC, MPC, multi-path communication, and hierarchical keying structures under a unified framework and  extended operational modes that capture deployment environments in which trusted key-management servers are intermittently available, replicated, thresholded, or entirely absent.
Such a systematization allows us to classify network designs according to their cryptographic assumptions and trust requirements. Using this framework, we analyze a broad range of architectural patterns: centralized Kerberos-like KDSs, replicated and threshold variants that mitigate single points of failure, hierarchical tree-based approaches that support scalability, MPC-enabled KDS clusters that enhance compromise resilience, and multi-path secret-sharing systems that eliminate trusted servers entirely. We complement this architectural taxonomy with a security 
evaluation that captures key properties relevant to PQ transition, including forward secrecy, post-compromise security, key compromise impersonation resilience, fault-tolerance, multi-client scalability, and network-layer adversary capabilities. By synthesizing results from PQC, symmetric cryptography, distributed systems, MPC, and network security, we provide an integrated view of how quantum-resistant key management can be achieved across diverse deployment environments. 

Our taxonomy is motivated by the observation that PQ transition pressures networks along several fundamentally different axes: \textcircled{1} the availability of cryptographic primitives, \textcircled{2} the trust and operational model of deployment environments, \textcircled{3} the key-distribution architecture, and \textcircled{4} the network-layer assumptions about path diversity and adversarial capabilities. Existing PQC transition efforts typically focus on a single axis—usually cryptographic substitution within a protocol stack—while real-world systems must evaluate multiple interacting dimensions simultaneously. 

We selected our taxonomy categories (crypto foundation, key-distribution architecture, trust model, key management lifecyle, network-layer integration, and security properties) because they jointly capture the essential design decisions that determine whether a network can remain quantum-safe under diverse deployment constraints. These dimensions also highlight environments where PQC is unavailable or impractical, where symmetric-only or multi-path constructions are required, and where hierarchical or MPC-backed KDS systems offer meaningful advantages. By organizing the design space along these axes, our systematization reveals not only what kinds of architectures are possible, but also which architectures are appropriate for specific operational contexts in the post-quantum era.

A recurring topic in PQ transition guidelines is \textbf{cryptographic agility}, often narrowly defined as the ability to replace cryptographic primitives within software stacks. While such algorithm-level agility is necessary, it is far from sufficient for networks. In practice, cryptographic agility in networks is a system-level property that depends on key-distribution architectures, trust assumptions, lifecycle management, protocol layering, and communication topology. A network system that can swap algorithms at endpoints may still lack the ability to rotate keys at scale, recover from compromise, adapt to heterogeneous trust environments, or operate when PQ-PKC is unavailable. This observation motivates our architectural and taxonomy-driven approach: rather than treating agility as a software feature, we consider it as an emergent property of network design. Viewed together, our taxonomy dimensions capture the structural enablers and limits of cryptographic agility in networks. Different combinations of cryptographic foundations, key-distribution architectures, trust models, lifecycle mechanisms, and deployment environments determine not only whether a system is quantum-resistant today, but whether it can adapt as cryptographic assumptions, standards, and threat models evolve over time. By organizing the design space along these axes, our taxonomy makes explicit which forms of agility are feasible in different operational contexts, and which are fundamentally constrained by architectural choices.

\begin{center}
\begin{tcolorbox}[
  enhanced,
  width=0.96\linewidth,
  colback=gray!10,
  colframe=black,
  arc=4pt,
  boxrule=0.8pt,
  left=6pt,
  right=6pt,
  top=6pt,
  bottom=6pt
]
\textbf{Cryptographic Agility Is a Network Property}

Cryptographic agility in networks extends far beyond algorithm substitution at endpoints. In networked systems, agility emerges from architectural choices spanning cryptographic foundations, key-distribution mechanisms, trust models, lifecycle management, protocol layering, and communication topology. A system may support multiple cryptographic algorithms and still lack the ability to rotate keys at scale, recover from compromise, or operate under partial PQ availability. This observation motivates a taxonomy-driven approach that treats cryptographic agility as a system design property rather than a software feature.

\end{tcolorbox}
\end{center}

This paper makes the following contributions:
\begin{enumerate}
\item A unified taxonomy of post-quantum network architectures spanning symmetric-only designs, PQ-PKI enabled systems, hybrid modes, and serverless multi-path constructions.
\item A systematization of centralized, distributed, hierarchical, MPC-backed, and multi-path key distribution mechanisms, analyzing their security assumptions, scalability properties, and suitability for PQ deployment.
\item A comparative security analysis that evaluates forward secrecy, key compromise tolerance, adversary models, side-channel considerations, and post-quantum threat resilience across these architectures.
\item A deployability analysis, discussing computational cost, network assumptions, multi-client support, and transition pathways for real-world systems.
\item
Challenges and open research directions for building quantum-resistant network security infrastructures, including opportunities at the intersection of PQC, MPC, zero-trust architectures, and distributed trust.
\item A set of best practices for PQ network design and migration, emphasizing architectural cryptographic agility, lifecycle-driven key management, and practical strategies for operating under heterogeneous trust models and partial PQ deployment.
\end{enumerate}

For convenience, Table~\ref{tab:acronyms} in Appendix~A summarizes the acronyms used throughout the paper.


%% file: relatedwork.tex
\section{Related Work}
A growing body of SoK and survey work 
analyzes the transition to post-quantum cryptography (PQC), but these efforts primarily focus on cryptographic primitives, implementation practices, or protocol-specific adaptations, leaving a gap in the understanding of PQ-resilient network architectures and key-distribution systems. Howe et al.’s SoK on PQ  cryptography design and implementation provides an extensive analysis of PQC algorithms, implementation pitfalls, and side-channel considerations, but is limited to the level of primitives rather than network architecture or key-management frameworks~\cite{howe2021sok}. Other similar works conduct an in-depth systematization of PQ and hybrid TLS designs. Alnahawi et al.’s SoK on PQ-TLS and their comprehensive 2024 survey analyze where and how to embed KEMs, how to construct hybrid handshakes, and the performance/soundness implications of different integration strategies~\cite{alnahawi2023pqtlseprint,alnahawi2024pqtlssurvey}. While these  analyses provide valuable guidance for secure transport-layer migration, they do not generalize their taxonomies to key-distribution infrastructures, symmetric-only architectures, or multi-path/threshold-based network primitives.

A second line of relevant research focuses on cryptographic agility and PQ transition frameworks. N{\"a}ther et al. present a systematic analysis of cryptographic agility concepts and design patterns, highlighting the challenges of algorithm substitution and phased transitions~\cite{naether2024agility}. Their complementary systematic literature review on PQ-migration techniques surveys engineering practices, developer approaches, and software-level considerations for PQC integration~\cite{naether2024slr}. While these works offer valuable insights for protocol evolution and software ecosystems, they do not analyze the architectural consequences of PQ transition for distributed key-distribution systems, nor do they address symmetric-only or trust-minimized environments where PQ-PKI may be unavailable.

Related domains contain additional SoK efforts—for example, surveys on PQ blockchain systems~\cite{mallick_quantum_blockchains_sok_2025}, hybrid PQC migration strategies, and unified quantum-protocol frameworks~\cite{kumibe2025hybrid}. These works provide insights on decentralized trust, hybrid crypto migration, and quantum-communication primitives, but they focus on specific application domains rather than the broader question of how to secure classical IP-based networks, Kerberos-style KDS deployments, or multi-path key-distribution systems in a PQ world.

Finally a closely related line of work focuses on quantum key distribution (QKD) networks and their key-management infrastructures. A recent and comprehensive survey by Dervisevic et al.~\cite{dervisevic2024qkd-survey} provides an in-depth systematization of key-management architectures for QKD-enabled networks. The authors analyze how quantum-generated keys are stored, refreshed, distributed, and synchronized across QKD nodes, with particular attention to trusted-node architectures, QKD control planes, and the interaction between quantum and classical management layers. While highly valuable, the scope of that survey is fundamentally different from ours. It assumes the availability of quantum communication hardware and QKD links, and focuses on the management of keys produced by quantum channels. In contrast, our work addresses PQ–resistant classical networks, where security must be achieved without relying on quantum communication, specialized hardware, or trusted QKD infrastructure. Our taxonomy spans symmetric-only, PQ-PKI, hybrid, threshold/MPC-based, and serverless multi-path architectures that operate over today’s IP-based networks and are deployable in environments where QKD is unavailable, impractical, or undesirable. As a result, the two surveys are complementary. The survey by Dervisevic et al.~\cite{dervisevic2024qkd-survey} systematizes how to manage quantum-generated keys once quantum networks exist, whereas our work systematizes how to design cryptographically agile, quantum-resilient key-distribution and management architectures for classical networks in the post-quantum era. Together, they provide a more complete picture of future secure networking across both quantum-assisted and purely classical infrastructures.

%% file: CryptographicFoundations.tex
\section{Taxonomy by Cryptographic Foundations}
A core dimension in evaluating PQ–resistant network architectures is the nature of the cryptographic foundations they rely upon. Different deployment environments, performance constraints, and availability of trust anchors lead to fundamentally different choices of primitives, which in turn influence the design of the entire key-management and authentication ecosystem. We classify cryptographic foundations into four categories: symmetric-only, PQ public-key, hybrid classical-PQC, and multi-path/information-theoretic approaches. This categorization enables a principled comparison across architectures with differing assumptions, security guarantees, and operational feasibility.

\subsection{Symmetric-Only Cryptography}

In some environments, such as constrained IoT deployments, high-assurance military systems, industrial networks, or legacy infrastructures—public key cryptography may be unavailable, impractical, or undesirable. These systems rely entirely on symmetric primitives (e.g., AES with sufficiently large keys, SHA-2/SHA-3 families, HMAC, pseudorandom number generator (PRNG)/quantum random number generator (QRNG) sources) and achieve authentication and key establishment through pre-shared secrets or symmetric trust anchors. These primitives include AES-GCM, AES-CTR, ChaCha20 with double-keying, hash-based authentication (HMAC, KDF chains), and PRNG/QRNG-based keying.
While secure against breaches due to Shor’s algorithm and only mildly affected by Grover’s algorithm, symmetric-only systems face challenges in scalability, forward secrecy, and distributed trust. Their security properties depend heavily on the protection and rotation of long-term symmetric keys and architectural decisions such as centralized KDSs, threshold servers, or multi-path secret sharing. This category corresponds to
the minimal cryptographic foundation for quantum-safe operation. Note that the size of the keys must be doubled - to protect again key breaches that leverage Grover’s algorithm. For very small devices this can be a problem and thus security must be based on architectural approaches taking into account the weaknesses of specific components of the network of interest.

\subsection{Post-Quantum Public-Key Cryptography (PQ-PKC)}

The second category encompasses systems that deploy PQ public-key primitives, including lattice-based KEMs (e.g., Kyber) and signatures (e.g., Dilithium, SPHINCS+) standardized by NIST. These primitives enable scalable key establishment, certificate-based authentication, and federated trust infrastructures analogous to today’s PKI. Networks operating under this foundation can maintain interoperability with modern protocols such as PQ-TLS, PQ-SSH, and PQ-IPsec. The inclusion of PQ-PKC expands the design space significantly, enabling decentralized trust, ephemeral keys for forward secrecy, and scalable cross-domain authentication. This category 
assumes availability of PQC libraries, certificate issuance, and secure key distribution via PQ-secure channels. 

PQ algorithms have distinct memory footprints and CPU cycle patterns, due to their complex operations, larger keys, and ciphertexts. These characteristics make them 
identifiable~\cite{pq_fingerprinting_2025} through classification techniques that analyze behavioral patterns and protocol-specific responses. Such classification can expose systems to risks, as attackers may exploit identifiable traits to launch denial-of-service, key recovery, or downgrade attacks.

\subsection{Hybrid Classical-PQC Foundations}

Many real-world deployments will undergo a long transition during which classical public-key primitives (e.g., ECDH, ECDSA) coexist with their PQ counterparts. Hybrid designs combine classical and PQ primitives either in parallel (e.g., dual-signature authentication) or compositionally (e.g., combining ECDH and PQ-KEM secrets via a KDF). Hybrid foundations provide a defense-in-depth approach: the system remains secure unless both the classical and PQ components fail. This foundation is currently the default in multiple standards bodies for migration-critical sectors because it mitigates risks of premature reliance on recently standardized PQC algorithms while avoiding the catastrophic failure associated with future quantum adversaries. It is however important to note that these approaches have increasead complexity and the cryptographic techniques that are combined must have ``comparable security'', making their analysis more complex.

\subsection{Multi-Path and Information-Theoretic Foundations}

A fourth category includes \emph{serverless or network-assisted key distribution mechanisms} based on multi-path secret sharing, randomized routing, or information-theoretic constructions. These systems may not require any public-key primitives and can operate even when PQC libraries or trust infrastructures are absent. Keys are derived by splitting entropy across multiple independent communication paths or mediators, so an adversary must compromise all paths to recover the secret. Such foundations benefit from information-theoretic secrecy properties and can be combined with symmetric KDF chains or QRNG sources to enhance forward secrecy and key freshness. These approaches are particularly relevant in environments with strong adversarial presence on the network (e.g., nation-state surveillance) or where trust-minimization is a requirement. Note that security of these approaches relies on network assumptions (e.g. observability and  interception) and not on computational assumptions.

\subsection{Summary of the Cryptographic Foundations Dimension}

This cryptographic-foundation taxonomy serves as the first axis of our classification framework, as it dictates the feasibility, scalability, and trust assumptions of any quantum-resistant architecture built atop it. 
Symmetric-only systems 
emphasize minimal assumptions and robustness but require architectural compensations such as MPC or multi-path routing. 
PQC-enabled systems 
support scalable and federated authentication. Hybrid systems simplify real-world migration, whereas information-theoretic multi-path systems address environments with strong adversarial capabilities or limited trust. 

\subsection{Research Gaps}
While this taxonomy captures the wide scope of cryptographic foundations used in quantum-resistant network design, it also shows several challenges that need to be addressed. Current research provides only partial guidance on how to construct scalable symmetric-only systems, how to rigorously analyze hybrid classical–PQC deployments, or how to integrate multi-path and MPC-based mechanisms into coherent security models. These gaps highlight the need for deeper theoretical, architectural, and empirical work to understand the strengths and limitations of each foundation in real-world settings. Specific gaps include:
\begin{itemize}
\item \emph{Lack of systematic design principles for symmetric-only networks.} Although symmetric-only quantum-safe designs are attractive for constrained or high-assurance environments, there is no unified theory for constructing scalable, forward-secure, and compromise-resilient symmetric-only networks. Open questions include: (i)  Achieving scalable key establishment without PKC. (ii) Identifying  best strategy to layer MPC, replication, or multi-path techniques to compensate for PKC absence. (iii) Formalizing the security guarantees of symmetric-only networks in adversarial routing environments.
\item \emph{Limited understanding of hybrid cryptographic foundations under dynamic migration.} Hybrid classical–PQC systems are widely suggested, but there is no analytical framework covering: (i) Optimal hybrid compositions (parallel vs. serial vs. KDF-composed). (ii) Long-term implications of partial trust in PQC or classical components. (iii) Migration strategies for large multi-vendor networks. (iv) Failure models for hybrid crypto (e.g., if PQ part weakens).
\item \emph{Absence of formal models for multi-path/information-theoretic key transport.} Multi-path secret-sharing mechanisms provide strong theoretical guarantees but lack: (i)  Formal adversary models combining network and cryptographic capabilities. (ii) Realistic analyses of path independence in mobile and cloud networks. (iii) Practical guidelines for rate-limiting, retransmission, and loss resilience. (iv) Integration frameworks with existing protocols such as KDS or PQ-TLS.
\item \emph{Limited analysis of cross-primitive combinations (e.g., PQC + multipath + MPC).} 
There is no work exploring joint designs - in particular their security guarantees and how to reconcile their different adversarial assumptions, although these combinations may provide stronger guarantees than any primitive alone.
\end{itemize}






%% file: KeyDistributionArchitectures.tex
\section{Taxonomy by Key Distribution and Establishment Architectures and Protocols}

Key distribution and establishment is a foundational problem in secure network design, and its relevance is amplified in the PQ setting. The choice of architecture determines not only the scalability and performance of the system, but also its resilience to compromise, its trust assumptions, and its ability to adapt to environments where PQC may or may not be available. We classify key-distribution architectures into five main categories: (1) centralized, (2) replicated and (threshold) distributed, (3) hierarchical, (4) contributory, and (5) serverless multi-path. Each category embodies distinct operational assumptions and security properties, and the taxonomy makes explicit the trade-offs that arise when deploying these architectures under different quantum-resistant cryptographic foundations.

\subsection{Centralized Key Distribution (Single/KDC-based)}
Kerberos-style Key Distribution Centers (KDCs) remain viable in a PQ threat model as long as (1) the core ticket protection stays symmetric and uses conservative parameters, and (2) any public-key add-ons (mainly for initial authentication and inter-realm trust) are upgraded to PQ primitives. The baseline Kerberos design uses symmetric session keys carried in tickets and protected by long-term symmetric keys at the KDC and services; quantum impact here is mainly Grover-style brute-force speedups, which are mitigated by using larger keys (e.g., AES-256-class) and avoiding low-entropy secrets~\cite{campagna2013kerberos}. In this setting, the PQ-safety is mainly decided by the operational rules:  (1) ensure that  key material is random/high-entropy (not password-derived), (2) keep ticket lifetimes and replay windows tight, and (3) use modern authenticated encryption/hashes with large security margins~\cite{baseri2024navigating,sherman2014needham}.

Public-key-supported Kerberos variants, such as PKINIT 
(RFC 4556~\cite{RFC4556}), are widely used because they avoid the need to provision per-user symmetric secrets and better align with enterprise certificate workflows. However, despite their improved scalability, they rely on conventional-secure primitives such as RSA/DH/ECDH. A required migration step is to replace those conventional-secure mechanisms with Module-Lattice (ML)-KEM, as standardized in FIPS 203~\cite{nist2024fips203}, and replace certificate-chain and KDC assertion signatures like ML-DSA, as standardized in FIPS 204~\cite{nist2024fips204}. Cross-realm authentication (RFC 6806 \cite{RFC6806}) has the same issue: any signature-based realm-to-realm validation requires a PQ signature suite, and any key establishment used in those paths should be a PQ-KEM or a hybrid construction if incremental rollout is required~\cite{rfc9794}. Practical experiments in Linux environments show that keeping the Kerberos message flow and ticket formats stable while upgrading only the crypto underneath is a realistic path, especially when the rest of the stack already supports crypto-agility policies~\cite{redhat2025pqc}.

One deployment hurdle for PQ-secure PKC-based Kerberos variants is the increased message size. LM-KEM ciphertexts and PQ signatures are larger than those of RSA/ECC variants, and Kerberos still often uses UDP for performance and legacy reasons. Larger AS-REQ/AS-REP payloads can exceed Maximum Transmission Unit (MTU), trigger fragmentation, and cause hard-to-debug drops; operationally, this pushes deployments toward TCP for Kerberos or toward carrying Kerberos exchanges over an HTTPS-based KDC proxy in constrained networks~\cite{mskkdcp}. In practice, the ``harvest now, decrypt later'' concern also changes prioritization: protecting session-key delivery early (e.g., the TGT/AS exchange) is a high-value target for PQ/hybrid key establishment, even if the full certificate ecosystem takes longer to transition~\cite{rfc9794}.

Beyond PQ-secure Kerberos variants, there are PQ-native designs that aim to reduce the single-point-of-failure risk of a centralized KDC or to improve key establishment properties. Lattice-based group-authenticated key exchange (GAKE) protocols can distribute the ``ticket-granting'' role across participants so that a group key is derived without a single online trusted server per session, while still providing authentication under lattice assumptions~\cite {zhang2025lattice}. Threshold Single Sign-On (SSO) schemes built from LWE, such as LPbT-SSO, explicitly replace a single KDC with multiple identity servers; no single compromised server can mint credentials or recover long-term secrets, which is closer to the availability/compromise model many Kerberos operators actually want~\cite{cao2025lpbt}. These approaches are not drop-in Kerberos replacements, but they are relevant when the KDC compromise model (or cross-realm trust sprawl) is the primary risk driver rather than just ``replace RSA/ECC'', as discussed further below.



\subsection{Fault-Tolerant and Distributed (Threshold) KDS}
To improve availability and operational robustness, a centralized KDS/KDC can be replicated across multiple servers. A common setup is hot-cold (active-passive) or active-active replication, where multiple replicas hold the same long-term key state and answer requests, with a replication protocol ensuring state convergence after failures. The main security issue is that replication expands the attack surface: once a node is compromised, the attacker can attempt to fork or roll back key state, or reply inconsistently to different clients (equivocation). The main engineering issue is keeping a strict, auditable ordering of key updates (e.g., key rotations, revocations, counters/epochs) so that tickets and derived keys remain verifiable and replay windows do not widen.

Byzantine fault-tolerant KDC clusters aim to keep the service correct even when some replicas are actively malicious. A standard approach is to run the KDC as a replicated state machine and require a quorum agreement on each state transition; the classical bound is \(n \ge 3f+1\) replicas to tolerate \(f\) Byzantine faults, with quorum certificates (e.g., \(2f+1\) matching votes) to commit operations. Practical Byzantine fault tolerance (PBFT)~\cite{castro1999pbft} gives the baseline structure for this, and the later ``proactive recovery'' work shows how to periodically refresh replicas to limit long-lived compromise and keep the system safe over time~\cite{castro2000proactive}. In a PQ-secure version, the BFT layer needs authentication that is not broken by Shor: either use symmetric-message authentication inside a closed replica group (pairwise MACs) or use PQ signatures, accepting that PQ signatures and public keys are larger than RSA/ECC and can increase replication bandwidth and latency~\cite{nist2024fips203,nist2024fips204,araki2016ht3pc}.

Replication and BFT reduce availability risk but do not remove the single point of trust: every replica still holds the full long-term secret unless extra mechanisms are used. Threshold KDS designs change this by splitting each long-term secret into \(n\) shares with a threshold \(t\), so no single node ever holds the complete key material; reconstruction or key-dependent operations require at least \(t\) shares. Shamir secret sharing is the standard building block, and it supports proactive refresh where shares rotate while the underlying secret remains fixed~\cite{shamir1979}. In Kerberos-like deployments, this can be applied at the ticket-issuing step by distributing the TGS/KDC secret across multiple servers; the PKDA line of work already treats Kerberos ticketing as a framework into which distributed cryptography can be inserted, shifting sensitive logic away from a single online KDC~\cite{chuang1997pkda}. For PQ-resilience, threshold designs are attractive because the core protection can remain symmetric-only (tickets encrypted/MACed under long-term symmetric secrets), while the distributed protocol prevents any single breach from revealing those secrets. COCA~\cite{coca_2002} also uses BFT with threshold cryptography, while verifiable secret sharing approaches allow for malicious detection.

MPC-backed KDS systems go further by having servers jointly compute keys/tickets without reconstructing secrets at any point, even transiently. The potential performance of MPC at the Kerberos scale is shown in~\cite{araki2016ht3pc} on high-throughput 3-party computation with an honest majority, which explicitly demonstrates Kerberos authentication/ticket-generation workloads using MPC throughput in the tens of thousands of queries per second. For stronger adversaries (fully malicious corruption), SPDZ-like protocols are suitable: they provide malicious security via authenticated secret sharing and a preprocessing phase, and they can be instantiated over rings that map well to blockcipher-based constructions~\cite{damgard2012spdz,cramer2018spdz2k}. Modern SPDZ-family work (e.g., MASCOT-style preprocessing) reduces the cost of malicious security, making threshold-KDC functions like PRF evaluation, ticket-field MAC computation, and policy checks realistic candidates for MPC execution~\cite{keller2016mascot}. DiSE (Distributed Symmetric-key Encryption) is directly relevant when the goal is to keep AES-class protection but eliminate single-node key custody: it formalizes distributed authenticated encryption where encryption/decryption is performed with threshold participation and security is defined against malicious subsets~\cite{agrawal2018dise}. This model aligns with a ``KDC as a key-usage oracle'' design, where the service never releases the long-term secret but only releases outputs derived from it (tickets, wrapped session keys, or key confirmations). In the PQ setting, these distributed and quantum-safe infrastructures (e.g.,~\cite{10063497}) remain meaningful even if public-key migration is blocked (e.g., policy constraints or legacy endpoints), because the security reduces to symmetric primitives plus the threshold PQ signature assumptions, not to the conventional-secure primitives~\cite{10.1145/3772274,agrawal2018dise}.

The trust model changes substantially when moving from single-server or replicated-server KDCs to threshold and Byzantine/MPC-based KDCs. A single-server KDC assumes one trusted machine holding the full master secret; compromise is total and immediate. Replication improves uptime but usually does not change that trust assumption: every replica is still a full-trust copy unless secrets are split. Threshold KDS assumes that fewer than \(t\) nodes collude or get compromised at the same time; this is a different (and typically stronger) risk model than ``no crash faults'' and also different from pure availability-driven replication. Byzantine/MPC-based KDCs further assume that even actively cheating nodes are possible and build correctness checks (quorums, MAC checks, consistency proofs) into the protocol itself, trading extra latency and coordination for a strictly smaller blast radius under compromise~\cite{castro1999pbft,cramer2018spdz2k,agrawal2018dise}.

\subsection{Hierarchical Key Distribution}
Kerberos-style systems and Logical Key Hierarchy (LKH) group key management share a common non-democratic trust model: a centralized infrastructure element (KDC or Group Controller) provisions symmetric secrets to scale authentication or secure multicast without requiring all principals to participate equally in key generation~\cite{butler2005kerberos-crossrealm,wallner1999rfc2627}. This centralized approach provides natural integration points for policy enforcement, auditing, and crypto-agility in enterprise and carrier-scale deployments where high membership churn makes pairwise keying operationally infeasible~\cite{wong2000keygraphs,harney1997gkmp}. LKH organizes members as leaves of a key tree, updating only keys along the affected path during join/leave events to achieve logarithmic rekey costs~\cite{wallner1999rfc2627,mcgrew2003oft}. One-Way Function Tree (OFT) optimizations further reduce bandwidth by enabling key derivation via hash functions rather than explicit transmission~\cite{mcgrew2003oft,aldarwbi2020keyshield}.

Hierarchical architectures partition trust and computation across organizational units or geographic regions, aligning key distribution with administrative boundaries as enterprises and infrastructure operators naturally segment their networks~\cite{dinker2025wsn,abdmeziem2025scada-taxonomy}. In 
PQ settings, these hierarchies integrate PQ-PKI to authenticate inter-domain relationships while maintaining lightweight symmetric operations for intra-domain updates~\cite{fsisac2025pqc,ghosh2024scada-multiphase}. However, symmetric-only hierarchies using inter-cluster symmetric keys face forward secrecy challenges, as compromise of long-lived inter-domain keys can cascade across multiple subtrees~\cite{aldarwbi2020keyshield,kumar2024pq-cloud}.

Kerberos cross-realm authentication extends this model through direct or transitive trust relationships anchored by shared inter-realm keys or PKI trust paths~\cite{butler2005kerberos-crossrealm,nist2024pqc-standards}. PQ Kerberos proposals leverage NIST-standardized ML-KEM for key agreement and ML-DSA for cross-realm authentication to protect ticket-granting infrastructure from quantum adversaries~\cite{nist2024pqc-standards,techrxiv2025pqc-zt,dervisevic2024qkd-survey}. Hybrid modes combining classical and PQ algorithms maintain backward compatibility during migration~\cite{nist2025crypto-agility,ietf2025composite-guidance,techrxiv2025pqc-zt}, with IETF proposals exploring KEM-based authentication mechanisms to provide quantum-resistant key exchange without sole reliance on vulnerable signature-based PKI channels~\cite{ietf2025kem-ikev2,ietf-tls-kdh}.

For PKI-backed authentication, Kerberos PKINIT enables certificate-based identity verification during initial KDC exchanges~\cite{redhat2025pqc-kerberos,adan2025quantum-auth}. Modern PKINIT implementations support PQ signatures such as ML-DSA in certificate chains, enabling quantum-resistant user and service authentication~\cite{redhat2025pqc-kerberos,arxiv2025applied-pqc-pki}. Hybrid and composite certificate authorities issuing dual-algorithm certificates (classical ECDSA/RSA combined with ML-DSA) facilitate gradual migration while preserving legacy client interoperability~\cite{arxiv2025applied-pqc-pki,ietf2025composite-guidance,thinkmind2024composite-certs}. This aligns with IETF and academic guidance emphasizing composite certificate frameworks and modular implementations for defense-in-depth during the PQ transition~\cite{ietf2025composite-guidance,arxiv2025applied-pqc-pki,techrxiv2025pqc-zt}.

The Iolus framework extends scalability through subgroup partitioning, where local Group Security Agents manage subgroup-specific rekeying to localize the impact of membership changes~\cite{mittra1997iolus,wong2000keygraphs}. Recent SCADA architectures like SKMA+/ASKMA+ combine LKH trees with Iolus-style subgrouping to align security boundaries with operational zones while maintaining centralized governance~\cite{abdmeziem2025scada-taxonomy,ghosh2024scada-multiphase}. Decentralized variants like DeGKM apply similar partitioning principles for content distribution networks~\cite{liu2024degkm,zhang2025iot-survey}.

Symmetric hierarchical schemes maintain distinct PQ advantages because core rekeying operations rely on AES and SHA~\cite{adan2025quantum-auth,aldarwbi2020keyshield}. Emerging SCADA and 5G/6G frameworks leverage this advantage through hybrid designs: heavy PQ signatures or KEMs protect inter-domain and bootstrapping channels, while intra-domain rekeying remains symmetric and hierarchical~\cite{ghosh2024scada-multiphase,smith2024_5g6g}. For IoT and automotive deployments, robust multicast authentication mechanisms complement confidentiality with origin authentication and replay protection, making hierarchical key structures central to both scalability and security in the post-quantum era~\cite{bella2025macsec,zhang2025iot-survey}.

\subsection{Contributory and Dynamic Key Establishment}

Unlike the non-democratic trust models of Kerberos-style KDCs and LKH-based group controllers, contributory group key establishment protocols enable all participants to jointly contribute to the group key without relying on a central trusted authority~\cite{kim2004tgdh-acm,amir2004robust-gka,nita2002cliques-dsn}. In basic Group Diffie-Hellman (GDH), each member contributes randomness that is cryptographically combined using modular exponentiation to produce a shared group key that no single party controls~\cite{kim2004tgdh-acm,amir2004robust-gka}. Nita-Rotaru's work on high-performance secure group communication established robust contributory key agreement protocols resilient to failures, partitions, and cascading membership events by integrating Group Diffie-Hellman with the Spread group communication system's Virtual Synchrony semantics~\cite{nita2001phd,amir2004robust-gka,nita2002cliques-dsn}. The Burmester-Desmedt (BD) protocol allows $n$ parties arranged in a logical cycle to establish a shared session key in a constant number of rounds, typically two, regardless of the group size~\cite{burmester1994secure}. In the first round, participants broadcast ephemeral public keys, and in the second, they compute and broadcast the ratio of the shared secrets formed with their left and right neighbors, allowing every member to derive the common group key~\cite{katz2003scalable}. The BD protocol is more suitable for static and high-latency WAN networks. Tree-Based Group Diffie-Hellman (TGDH) organizes members into a key tree structure where intermediate nodes represent partial keys computed via pairwise Diffie-Hellman exchanges, enabling logarithmic-cost dynamic rekeying upon member joins and leaves~\cite{kim2004tgdh-acm,lee2002tgdh-icnp}. These contributory protocols are particularly well-suited for dynamic next-generation networked environments such as mobile ad-hoc networks, vehicular platoons, and tactical communication groups (e.g., \cite{YavuzAlagozAnarim2010}) where establishing a priori trust relationships with a centralized controller is impractical or undesirable~\cite{buttyan2012invitation-tgdh}. However, state-of-the-art GDH and TGDH implementations predominantly rely on elliptic-curve Diffie-Hellman (ECDH), which Shor's algorithm renders vulnerable to quantum attacks~\cite{nist2024pqc-standards}. 

The fundamental challenge in PQ contributory key establishment is the absence of efficient Diffie-Hellman-like key exchange or cryptographic pairing operations in lattice-based, code-based, and most other PQ cryptographic families~\cite{ietf2025pqc-engineers}. While NIST-standardized ML-KEM provides key encapsulation, it is inherently asymmetric and non-commutative, preventing direct adaptation of classical GDH protocols that rely on the commutativity property. Isogeny-based constructions like CSIDH initially appeared promising due to their support for commutative group actions analogous to Diffie-Hellman~\cite{castryck2018csidh,kuleuven2018csidh}, but SIDH (the predecessor non-commutative variant) was cryptanalytically broken in 2022, and CSIDH itself suffers from prohibitively slow performance and ongoing security scrutiny that make it impractical for dynamic group settings~\cite{castryck2018csidh}. Consequently, achieving efficient PQ contributory group key establishment requires fundamentally different cryptographic primitives and protocol structures.

Recent efforts have focused on lattice-based group key exchange protocols that achieve contributory properties through compiler-based constructions from post-quantum KEMs. The works of Cantos et al.~\cite{pablos2020kyber-gake,pablos2022iet-kyber} proposed the first compiled construction toward PQ group key exchange using Kyber. It also transforms a two-party KEM into a multi-party group key agreement protocol by organizing participants into a tree structure where each node performs a KEM encapsulation/decapsulation sequence, ultimately deriving a shared group key without requiring Diffie-Hellman commutativity~\cite{pablos2020kyber-gake,pablos2022iet-kyber}. Subsequent work has extended these ideas to dynamic lattice-based authenticated group key exchange (AGKA) protocols based on Learning With Errors (LWE) or Ring-LWE problems~\cite{zhang2023lattice-gka-tworounds,li2024lattice-dgka}. These protocols typically employ a coordinator or sequential message-passing structure to compute intermediate shared values, though this introduces additional rounds compared to classical TGDH~\cite{zhang2023lattice-gka-tworounds}.

Practical deployment of PQ group key establishment has been explored in several critical application domains where dynamic membership and quantum resilience are essential. In vehicular networks, module lattice-based schemes combining Kyber and Saber have enabled authenticated key exchange with low handshake latency, suitable for high-mobility V2X environments, demonstrating fast key generation and verification times necessary for platoon coordination and collision avoidance~\cite{zhang2025v2x-lightweight,yadav2024blockchain-vanet,wu2025pq-vanet-mac}. For unmanned aerial vehicles (UAVs), PQC aims to address the unique challenges of resource-constrained drone swarms operating over bandwidth-limited wireless channels. Some implementations integrate Kyber and Dilithium to achieve encapsulation and signature verification per message for authentication in drone identification, friend-or-foe (IFF) systems~\cite{demir2024uav-pqc-survey,alhaj2025uav-mavlink-pqc,scidir2024uav-swarm-survey}. Beyond algorithmic solutions, physical-layer QKD offers an alternative path for secure group key establishment, utilizing quantum conference protocols where a trusted sponsor distributes quantum states to participants to form a shared key~\cite{nature2025heterogeneous-qkn,nict2025qkd-multiplex}. While QKD requires specialized hardware and relays, experimental heterogeneous networks have demonstrated its viability for some distributed trust services, like quantum Byzantine agreement, alongside PQC to create robust hybrid architectures~\cite{nature2025heterogeneous-qkn,vasco2025scalable-qkd-hybrid}.

\subsection{Multi-Path and Secret-Sharing Solutions}
At the other end of the design spectrum are serverless mechanisms, where key establishment relies on multi-path communication and secret sharing rather than trusted servers. A session key is derived by splitting entropy across multiple independent network paths, forcing an adversary to compromise all paths simultaneously to recover the secret. These architectures are attractive for high-threat environments, zero-trust deployments, or scenarios where PQC is unavailable. They also enable strong forms of forward secrecy and information-theoretic protection. However, they depend heavily on network topology assumptions, path independence, path-failure handling, and reliable entropy aggregation. Integrating multi-path systems with centralized or distributed KDSs is an emerging hybrid direction that promises improved entropy quality and robustness but remains theoretically underdeveloped.

Specifically, a class of these protocols
leverages the concept of breaking a message into shares and
sending each share over a randomly chosen path, possibly
changing with time. For example, 
Lou and Fang~\cite{LouFang2001Multipath} propose
a secret sharing scheme over multiple network paths to provide
confidentiality with significantly lower computational requirements. Dolev and Tzur-David~\cite{DolevTzurDavid2017SecureInterconnection} propose a secret sharing
scheme over multiple network paths to mitigate the problem
of stolen or short keys.
Ahmadi et al.~\cite{AhmadiSafaviNaini2014Multipath} and Safavi-Naini et al.~\cite{SafaviNainiPoostindouzLisy2017PathHopping} considered
models where shares of the message are sent over multiple
paths that are changing (switching) in each time interval, and
proved that the system provides information theoretic security
and so stays secure against a quantum computer. However, 
such schemes rely on the assumptions that paths are atomic and
packets travel on such paths with the same delay. That is not the case
in real networks where paths have multiple hops, and each
hop (and path) can have different delays. Rashidi et al.~\cite{mpss_ndss_2021} showed that 
all schemes whose security relies on transfer
atomicity and all paths having the same delay, suffer from a 
side-channel attack, referred to as Network Data Remanence (NDR).

Many serverless schemes have been proposed in the context of sensor networks. Newell et al.~\cite{node_capture_2014} analyzed key establishment protocols for wireless sensor networks (WSNs) with a special focus on resilience to node capture—a major threat in sensor networks because physical capture of nodes exposes all keys stored on them.
They propose a new multi-path protocol tailored for sensor networks that uses an encoding scheme to improve resilience to node capture.


\input{table4}   

\subsection{Summary of Key Distribution Dimension}

\begin{tcolorbox}[
  enhanced,
  width=0.96\linewidth,
  colback=gray!10,
  colframe=black,
  arc=4pt,
  boxrule=0.8pt,
  left=6pt,
  right=6pt,
  top=6pt,
  bottom=6pt
]
\textbf{Design Trade-Offs in PQ-Resistant Key Distribution}
\begin{itemize}
\item  {\bf Centralized KDS:}
Efficient and familiar, but fragile under compromise and difficult to harden for PFS
\item {\bf Fault-Tolerant and Distributed (Threshold) KDS:} Enhances availability and compromise resiliency, but at the cost of coordination, latency and recovery complexity. 
\item {\bf Hierarchical KDS:} Scales well but amplifies compromise propagation risks
\item {\bf Contributory Key Establishment:} Enables dynamic and joint key creation in mobile systems
\item {\bf Serverless Multi-Path:} Minimal trust assumptions but dependent on topology and path diversity
\end{itemize}

\end{tcolorbox}

This taxonomy shows that no single key distribution architecture dominates across all deployment scenarios (see Table 1). Centralized and hierarchical KDSs offer operational familiarity and efficiency; replicated and threshold-based designs improve availability and compromise resilience; and serverless multi-path constructions offer strong security under minimal trust assumptions. When combined with the cryptographic-foundation taxonomy, these architectural classes expose critical design trade-offs: which architectures remain viable in symmetric-only settings, which ones benefit most from PQ-PKI, and which ones support scalable and resilient deployment in adversarial environments.

\subsection{Research Gaps}
Each of the previous approaches has specific research gaps that we discuss in what follows.

\textbf{Gaps in Centralized KDS Designs}
\begin{itemize} 
\item \emph{Missing frameworks for PQ hardening of classical KDS protocols.} Classical Kerberos and similar centralized systems have received only partial attention in the PQ context. There is no comprehensive design framework for: (i) Replacing (or avoiding) public-key pre-authentication in purely symmetric 
settings; (ii) Ensuring forward secrecy without PKC; (iii) Robust rotation for long-term symmetric keys at scale.
\item \emph{Lack of compromise recovery mechanisms.}
If a centralized KDS is compromised, recovery procedures are extremely costly. Currently, there are no standardized or formally analyzed recovery workflows (e.g., post-compromise rekeying, key rebinding, salt rotation) for PQ-era KDS deployments.
\end{itemize}

\textbf{Gaps in Fault-Tolerant and Distributed KDS}
\begin{itemize}

\item \emph{Absence of PQ-secure replication and state-synchronization protocols.}
Replication drastically changes the threat surface. Yet: (i)
No PQ-safe BFT replication protocols have been tailored to KDS workloads; (ii)  No state-merging protocols consider PQ adversaries who can break classical signatures; (iii) Cross-datacenter synchronization under PQ constraints is unexplored.

\item \emph{Lack of analyses of consistency–security trade-offs} Replication introduces consistency models (eventual, strong, BFT). However, the security costs of these choices in PQ or symmetric-only environments remain unquantified.

\item \emph{Absence of end-to-end formal models combining MPC and key distribution.} Existing MPC work focuses on specific subroutines, not full KDS workflows. Missing elements include: (i)  Composition proofs for ``MPC + symmetric KDF + session key distribution;'' (ii) Adversary models combining server compromise and partial network manipulation; (iii)
    Analyses of side channels and leakage during MPC-based ticket generation.

\item \emph{Lack of performance and scalability benchmarks.}
There is no comprehensive empirical evaluation answering questions such as: (i)  How many MPC servers are needed for a global-scale KDS? (ii)  What is the latency overhead for (t,n) threshold KDS under PQ assumptions? (iii)
    What are the minimal trust thresholds required for resilience to quantum adversaries?

\item \emph{No proactive-security designs tailored for PQ-era threat models.}
MPC-backed KDS could benefit from proactive resharing, key refresh, and compromise recovery, but proactive PQ-era KDS protocols do not exist.

\end{itemize}

\textbf{Gaps in Hierarchical Key-Distribution Systems}
\begin{itemize}
\item \emph{Lack of analyses concerning compromise propagation in PQ and symmetric-only hierarchies.}
A breach in a high-level node (regional KDS or domain anchor) may compromise thousands of subdomains.
There is no formal model for: (i) Compromise propagation in hierarchical symmetric-only architectures; (ii)
Thresholding strategies that contain damage across layers,
PQ-safe inter-domain trust with controlled blast radius.

\item \emph{Lack of deployment-validated hierarchical PQ architectures}
Although theory suggests that hierarchies should scale well, empirical evidence is lacking. There are no real-world or simulated studies quantify latency, churn, or fault behavior of PQ-enabled hierarchical KDSs. The behavior of PQ-signature-based inter-domain authentication under heavy load is unknown.
\end{itemize}

\textbf{Gaps in Contributory and Dynamic Key Establishment}
\begin{itemize}

\item \emph{Absence of native commutative primitives for contributory PQC}. Unlike the classical ECDH used in GDH and TGDH, standardized PQC algorithms (e.g., ML-KEM) lack the commutative property. While isogeny-based schemes (e.g., CSIDH) offer commutativity, they remain impractical due to prohibitive performance bottlenecks or security vulnerabilities (as seen with SIDH). Consequently, researchers are forced to rely on ``compiler'' approaches that construct group keys via sequential or tree-based KEM encapsulations. There is a distinct lack of native, efficient, and standard-compliant PQC primitives that enable single-round or constant-round contributory key agreement without the high communication overhead inherent in KEM-based constructions.

\item \emph{Inefficiency in dynamic rekeying and certificateless group establishment}. Existing PQC group key agreement proposals are ill-suited for the dynamic requirements of high-churn environments, such as UAV swarms or vehicular platoons. The large bandwidth requirements of lattice-based keys make rekeying operations prohibitively expensive when members frequently join or leave. Furthermore, there is a scarcity of efficient Certificateless Authenticated Group Key Agreement (CL-AGKA) schemes in the post-quantum domain. Current solutions often rely on heavy PKI management, which contradicts the ad-hoc nature of mobile networks, while the few existing certificateless PQC proposals suffer from massive key sizes that are unsuitable for resource-constrained edge devices.

\item \emph{Lack of fault-tolerant and partition-resilient PQC group protocols}. While classical literature successfully integrated cryptographic primitives with distributed system semantics (e.g., Virtual Synchrony) to handle network partitions and cascading membership events, this ``systems-level'' resilience is largely absent in PQC research. Current studies focus almost exclusively on cryptographic correctness and secrecy under ideal network conditions. There is no formal modeling or empirical analysis regarding how heavy PQC payloads interact with lossy networks, or how contributory PQC protocols recover from participant crashes and message drops during the complex, multi-round exchanges required by KEM-based trees.

\end{itemize}

\textbf{Gaps in Serverless Multi-Path Key Distribution}
\begin{itemize}
\item \emph{Lack of realistic models of path independence and adversary capabilities.} 
Multi-path key distribution offers strong theoretical guarantees. However: (i) Path-independence assumptions rarely hold in mobile networks, SDN, or cloud providers; (ii) No empirical analyses document real-world adversary coverage across autonomous systems (ASs)\footnote{An AS is collection of IP networks and routers under the control of a single administrative entity that presents a common routing policy to the Internet, cellular paths, data centers, or overlay routes.}. Adversaries may control large fractions of the network topology, including Internet Service Providers or ASs, thereby undermining assumptions of path independence in multi-path key-distribution schemes.

\item \emph{Unclear failure and loss-recovery semantics.}
Practical deployments require: (i) Packet-loss tolerance; (ii)
    Reconstitution of partial shares; (iii) 
    Anti-correlation techniques to avoid cross-path timing leakage.
    These two have not yet been much investigated.

\item \emph{There is no hybrid architectures combining KDS + multi-path entropy.}
A promising but unexplored direction is combining a KDS (centralized or hierarchical) with multi-path supplementing entropy, forward secrecy, or post-compromise recovery
There is no design or analysis of such hybrid systems—even though they could offer superior security in high-value settings.
\end{itemize}

\textbf{Cross-Cutting Architectural Gaps}
\begin{itemize}
\item \emph{Lack of comparative analysis across the architectural spectrum.}
No work compares symmetric-only KDS, hierarchical PQ-PKI KDS, MPC-backed KDS, and multi-path systems under: (i)  Uniform adversary models, (ii)  Common performance baselines, (iii)
    Realistic deployment scenarios (enterprise, IoT, NextG networks).

\item \emph{Absence of transition roadmaps from classical to PQ-safe architectures.}
While protocol-level migration strategies exist, architecture-level transitions have no guidance.
Questions remain unanswered: (i) How does a classical Kerberos realm migrate to PQ-Kerberos or MPC-Kerberos? (ii)
    How should organizations evolve from centralized to hierarchical or threshold-based KDSs? (iii)
    When should symmetric-only fallback modes 
    be maintained?
    \end{itemize}

%% file: table4.tex
\begin{table*}[t]
\centering
\footnotesize
\caption{Comparison of Key Distribution Architectures in PQ–Resistant Networks}
\label{tab:kds-comparison}
\begin{tabular}{|p{3.2cm}|p{2.6cm}|p{1.7cm}|p{1.1cm}|p{2.6cm}|p{1.1cm}|p{1.8cm}|}
\hline
\textbf{Architecture} &
\textbf{Trust Assumption} &
\textbf{PQ-PKC Required} &
\textbf{PFS} &
\textbf{Post-Compromise Recovery} &
\textbf{Scalability} &
\textbf{Operational Complexity} \\
\hline

Centralized KDS / KDC
(Kerberos-like) &
Fully trusted single authority &
No &
Weak &
Poor (global rekey often required) &
High &
Low \\
\hline

Replicated / BFT KDS &
Trusted replicas; fault tolerance &
No &
Weak--Medium &
Limited (replica compromise propagation) &
High &
Medium \\
\hline

Hierarchical KDS
(domain-based) &
Trusted domain anchors &
Often &
Medium &
Limited (domain-level recovery) &
Very High &
Medium \\
\hline

Threshold / MPC-based KDS &
Partial trust ($t,n$); no single point &
No &
Strong &
Strong (proactive resharing, healing) &
Medium &
High \\
\hline

Contributory Group Key Establishment &
No trusted server; all members contribute &
Yes &
Strong &
Medium (rekeying required) &
Medium &
High \\
\hline

Serverless Multi-Path Key Distribution &
Trust-minimized; path diversity assumptions &
No &
Strong &
Strong (fresh entropy injection) &
Low--Medium &
High \\
\hline

Server-Assisted Multi-Path Key Distribution & Partially trusted server for path selection; key material never fully reconstructed & No & Strong &Strong (path re-randomization; fresh entropy injection)& Medium & High\\
\hline

\end{tabular}
\end{table*}


%% file: TrustThreatModel.tex
\section{Taxonomy by Trust and Threat Model}
Beyond cryptographic primitives and key-distribution architectures, PQ–resistant network designs are fundamentally shaped by their trust assumptions and threat models. In practice, networks differ widely in how much trust they place in servers, intermediaries, administrators, and the network itself, as well as in the capabilities they attribute to adversaries. A systematic taxonomy must therefore make these assumptions explicit. We classify PQ network designs along four trust-and-threat categories: fully trusted authority models, threshold-trust models, trust-minimized or serverless models, and zero-trust adversarial network models. This taxonomy clarifies which architectures remain secure under partial compromise, insider threats, or powerful network-level adversaries, including those equipped with quantum computational capabilities.

\begin{center}
\begin{tcolorbox}[
  enhanced,
  width=0.96\linewidth,
  colback=gray!10,
  colframe=black,
  arc=4pt,
  boxrule=0.8pt,
  left=6pt,
  right=6pt,
  top=6pt,
  bottom=6pt
]
\textbf{Threat Model Assumptions in PQ Networks}

Throughout this paper, we assume adversaries with (i) quantum computational capabilities sufficient to break classical PKC, (ii) long-term traffic recording capabilities, and (iii) partial control over network infrastructure, including routing and key-management components. Our taxonomy makes explicit how different network architectures tolerate these threats, rather than assuming a uniform adversary model across deployments.

\end{tcolorbox}
\end{center}


\subsection{Trusted Server / Trusted Authority Model}
In the fully trusted authority model, one or more centralized entities—such as a KDS, Key KDC, or CA, are assumed to be trustworthy, uncompromised, and continuously available. Clients rely on these authorities for authentication, authorization, and session-key establishment. Classical Kerberos deployments and PKI-based authentication systems fall into this category. In the PQ context, such models can operate either in symmetric-only settings 
or with PQC-enabled PKI. While this trust model simplifies system design and enables efficient key distribution, it introduces strong assumptions: compromise of the trusted authority can lead to system-wide failure, mass impersonation, or long-term key exposure. Consequently, fully trusted models are most appropriate in tightly controlled environments, but they motivate the need for replication, auditing, and cryptographic hardening in PQ deployments.

\subsection{Threshold-Trust Model}
Threshold-trust models relax the assumption of a single fully trusted authority by distributing trust across multiple servers or administrators. In these systems, no single entity possesses sufficient information to compromise security; instead, correctness and confidentiality are guaranteed as long as fewer than a threshold number of components are compromised. Examples include threshold Kerberos variants, MPC-backed KDS architectures, and distributed CA services.

In the PQ setting, threshold models are particularly attractive because they reduce the impact of both classical and quantum-enabled attackers who may target key-management infrastructure. These models support compromise resilience, proactive key refresh, and gradual recovery from partial breaches. However, they introduce new challenges, including increased latency, coordination overhead, complex failure handling, and the need for formal composability guarantees under realistic adversary models.

\subsection{Trust-Reduced / Trust-Minimized Models}
Trust-minimized models aim to reduce or eliminate reliance on trusted servers altogether. Instead of assuming the existence of a KDS or CA, these systems derive security from cryptographic techniques such as multi-path secret sharing, distributed entropy aggregation, or information-theoretic constructions. In such designs, session keys are established by splitting secrets across multiple independent network paths or mediators, ensuring that no single compromised component reveals sufficient information.

These models are well-suited to high-threat environments, cross-organizational settings, or deployments where centralized trust is undesirable or infeasible. They are particularly relevant in 
scenarios where PQ-PKI is unavailable, and in adversarial network environments where server compromise is likely. Their security, however, depends heavily on assumptions about network path independence, adversary coverage, and traffic analysis resistance—assumptions that are difficult to validate in practice.
\subsection{Zero-Trust Network Assumptions}
Zero-trust models assume that no network component—internal or external—should be implicitly trusted~\cite{Katsis}. Every entity must continuously authenticate, and compromise of any subset of components is treated as a realistic possibility. In PQ settings, zero-trust assumptions extend to adversaries capable of breaking classical public-key cryptography, observing network traffic at scale, and selectively compromising key-management servers or routing infrastructure.

Under this model, architectures must tolerate insider threats, partial infrastructure compromise, and active network manipulation, including man-in-the-middle attacks and path interception. Designs combining symmetric cryptography, frequent key rotation, threshold trust, and multi-path communication are particularly relevant here. Zero-trust models thus serve as a unifying threat model that motivates hybrid architectures combining PQC, MPC, and serverless techniques to achieve defense-in-depth against quantum-enabled adversaries.

\subsection{Leveled Security Assumptions}
There are multiple levels of security assumptions for quantum resistance.  Different levels of threat assumptions are as follows:  (1) CRQC is not known to exist, no PQC adopted, subjected to harvest-now-decrypt-later (HNDL) threats, (2)   the entity has PQC deployed but only for certain scenarios but not for all threat scenarios thus subjected to HNDL, (3) PQC is widely adopted but has been found to be untrusted, may have vulnerabilities due to misconfigurations or incorrect implementations, (4) PQC is widely adopted but legacy systems are unable to adopt or are hard to be phased out, (5) PQC is widely adopted but other security and compliance changes and assumptions are making it harder to maintain quantum resistance with the current deployments 

\subsection{Summary of Trust and Threat Dimension}
This taxonomy highlights that trust assumptions are not binary but exist along a spectrum, from fully trusted centralized authorities to trust-minimized and zero-trust designs. Each trust model interacts closely with the chosen cryptographic foundation and key-distribution architecture, shaping the achievable security properties and deployment trade-offs. Making trust and threat assumptions explicit is therefore essential for evaluating the robustness of post-quantum network designs and for guiding the selection of appropriate architectures in diverse operational environments.

\subsection{Research Gaps}
While the taxonomy clarifies the spectrum of trust assumptions and adversarial capabilities underlying PQ–resistant network designs, it also exposes several important research gaps. These gaps reflect the fact that trust models in current systems are often implicit, under-specified, or misaligned with realistic post-quantum threat scenarios.

\textbf{Gaps in Fully Trusted Authority Models}
\begin{itemize}
\item \emph{Insufficient PQ threat modeling for trusted key-management servers.}
Most existing analyses of centralized KDSs or CAs assume classical adversaries and do not adequately capture post-quantum risks. In particular, there is limited work on: (i)
Modeling the long-term impact of retrospective decryption (``harvest now, decrypt later'') on centralized authorities; (ii) Quantifying the consequences of delayed compromise discovery in PQ settings; (iii) Understanding how trust in a central authority should evolve as PQ capabilities emerge incrementally.

\item \emph{Lack of formal recovery models after authority compromise.} Once a trusted authority is compromised, current systems provide little guidance on secure recovery. There are no standardized post-quantum models for: (i)  Re-establishing trust after a KDS or CA breach; (ii) Limiting the blast radius of compromised long-term secrets. (iii) Supporting secure re-enrollment of clients without relying on already-compromised trust anchors.
\end{itemize}

\textbf{Gaps in Threshold and Partial-Trust Models}
\begin{itemize}
\item \emph{Absence of unified adversary models for threshold trust.} Threshold and MPC-based systems often assume a fixed bound on the number of corrupt servers, but they have several shortcomings: (i)  Do not model adaptive adversaries who compromise nodes over time; (ii) Rarely consider combined network-level and server-level compromise; (iii)
Lack specific adversary definitions that combine cryptanalytic power with infrastructure control.

\item \emph{Limited composability guarantees across trust layers.}
Threshold trust is typically analyzed in isolation. However results are missing on several aspects: (i)  End-to-end composability results when threshold KDSs are combined with hierarchical architectures or PQ-PKI; (ii) Clear guidance on how threshold trust interacts with key rotation, delegation, and cross-domain authentication; (iii) Formal treatments of mixed-trust environments (e.g., threshold at the core, centralized at the edge).
\end{itemize}

\textbf{Gaps in Trust-Minimized and Serverless Models}
\begin{itemize}
\item \emph{Unrealistic or unvalidated trust assumptions about the network.}
Serverless and multi-path approaches often assume independent paths or non-colluding intermediaries. However there are several issues: (i) Real-world networks exhibit path correlation due to routing policies, shared infrastructure, and centralized providers; (ii) Adversaries may control large fractions of the network topology (e.g., ISPs, cloud backbones); (iii) There is no systematic methodology for validating path independence assumptions in practice.

\item \emph{Lack of formal threat models for combined network and cryptographic attacks.} Current analyses rarely capture adversaries that: (i) Observe traffic timing and metadata across multiple paths; (ii)  Selectively drop, delay, or replay secret shares; (iii) Combine quantum cryptanalysis with network-layer manipulation. A unified threat model encompassing these capabilities remains an open problem.
\end{itemize}

\textbf{Gaps in Zero-Trust and Highly Adversarial Models}
\begin{itemize}
\item \emph{No consensus definition of ``zero trust'' in the PQ context.}
While zero-trust networking is widely discussed, its application to PQ security remains informal. Open questions include: (i)  What trust assumptions are realistic once classical PKC is broken? (ii) How frequently must identities and keys be revalidated? (iii) Which components, if any, can be trusted transiently or probabilistically?

\item \emph{Insufficient integration of insider threats and long-lived 
adversaries.}
PQ adversaries may operate over long time horizons, slowly accumulating partial information. Existing models do not adequately address: (i) Insider threats combined with quantum decryption capabilities; (ii)  Long-term leakage from side channels or operational metadata; (iii) Gradual erosion of trust rather than instantaneous compromise.
\end{itemize}

\textbf{Cross-Cutting Gaps Across Trust Models}
\begin{itemize}
\item \emph{Lack of comparative evaluations under uniform threat assumptions.}
There is no systematic comparison of trust models under a common PQ adversary framework. As a result: (i) Claims about resilience, compromise tolerance, and robustness are difficult to compare; (ii)  Design decisions are often based on intuition rather than evidence; (iii) Trade-offs between trust centralization, performance, and security remain poorly understood.

\item \emph{Absence of trust-transition models during PQ migration.}
Real-world systems will transition between trust models over time (e.g., centralized to threshold to hybrid). However: (i)  No formal models describe how trust assumptions should evolve during PQ migration; (ii)  There is little guidance on safely operating mixed trust models concurrently; (iii) The security risks of transitional phases are largely unexplored.
\end{itemize}

The discussion on those gaps shows that trust assumptions in PQ network security are often implicit, static, and insufficiently aligned with realistic adversary capabilities. Addressing these gaps will require new threat models that combine quantum cryptanalysis, insider compromise, and network-level adversarial control, as well as new architectural designs that support dynamic trust, compromise recovery, and long-term resilience.

%% file: KeyManagementLifecyle.tex
\section{Taxonomy by Key Management Lifecyle}
While cryptographic foundations, key-distribution architectures, and trust models determine how keys are established, long-term security in PQ networks critically depends on how keys are managed over their entire lifecycle. In practice, many failures in cryptographic systems are due not to weak algorithms but from improper key generation, excessive key lifetimes, poor rotation practices, or inadequate compromise recovery. PQ readiness thus requires a holistic view of key management lifecycle processes, rather than isolated algorithm replacement. We then classify quantum-resistant network designs according to how they address the stages of the key lifecycle: key generation, key distribution, key usage, key rotation and revocation, key storage and protection, and post-compromise recovery.

\subsection{Key Generation}
Key generation determines the initial entropy and security margin of all subsequent cryptographic operations. In PQ networks, symmetric keys must be generated with sufficient entropy to withstand Grover’s algorithm, while PQC systems require high-quality randomness for lattice-based or hash-based constructions. It is critical that robust entropy sources be available and that one should not rely on legacy randomness assumptions during PQ migration. In practice, one may rely on software PRNGs, hardware random number generators, or quantum random number generators (QRNGs). Architectures that incorporate distributed or multi-path entropy sources can further strengthen resilience against biased or compromised randomness, but introduce additional complexity in entropy aggregation and validation.
\subsection{Key Distribution}
Key distribution governs how freshly generated keys are securely delivered to communicating parties. Lifecycle-aware systems distinguish between one-time session keys, short-lived derived keys, and long-term master secrets. It is critical that PQ transitions must explicitly account for how keys are established, refreshed, and replaced over time, especially in environments vulnerable to ``harvest-now, decrypt-later'' attacks. In symmetric-only settings, this typically involves KDS-mediated distribution or multi-path secret sharing, while PQ-enabled systems may rely on KEM-based key establishment. Lifecycle-aware architectures often combine these mechanisms to reduce key exposure windows and improve forward secrecy.

\subsection{Key Usage and Scope Control}
Once established, keys are used for specific cryptographic purposes—encryption, authentication, integrity, or derivation of subordinate keys. A lifecycle-oriented taxonomy distinguishes architectures that enforce strict key separation and purpose limitation from those that reuse keys across functions or sessions. It is critical to limit excessive key reuse in PQ systems, as longer key lifetimes increase exposure to future cryptanalytic advances. Lifecycle-aware designs therefore emphasize narrowly scoped keys, session-bound usage, and hierarchical derivation via KDFs to limit the impact of key compromise.
\subsection{Key Rotation, Expiration, and Revocation}
Key rotation is a critical element of PQ resilience. NIST SP 800-208 emphasizes that PQ readiness requires shorter cryptoperiods, proactive rotation, and explicit revocation mechanisms, especially for long-term keys that may be exposed today but exploited in the future. In symmetric-only architectures, rotation is often implemented via hash chains, epoch-based rekeying, or periodic KDS-assisted updates. In PQ-PKI systems, rotation and revocation rely on certificate lifetimes, revocation lists, or automated renewal protocols. Architectures that lack efficient rotation mechanisms are particularly vulnerable to long-term compromise in post-quantum threat models.

\subsection{Key Storage and Protection}
Secure key storage is a frequently underestimated aspect of the lifecycle. Keys may reside on clients, servers, or distributed across multiple nodes in threshold systems. It is important to protect keys both at rest and in use, using hardware isolation (e.g., Hardware Security Modules (HSMs) or Trusted Execution Environments (TEEs)) where appropriate, However it is important to note that such components introduce additional trust and supply-chain assumptions. Distributed and MPC-based architectures address storage risks by ensuring that no single component ever holds a complete secret, thereby reducing the impact of server compromise. Lifecycle-aware taxonomies must therefore distinguish between centralized, hardware-protected, and distributed key-storage models.

\subsubsection{HSM- and TEE-Based Key Storage}
HSMs and TEEs provide hardware-isolated key storage and cryptographic execution, preventing raw keys from being exposed to the host operating system or application memory. In this model, long-term symmetric keys, PQ private keys, or KDS master secrets are generated and stored inside protected hardware boundaries, and cryptographic operations are performed internally. From a PQ perspective, HSMs and TEEs offer strong protection against software compromise and memory disclosure attacks, which is particularly valuable for centralized KDSs, certificate authorities, or PQ-PKI services. However, these mechanisms introduce new trust assumptions: reliance on vendor correctness, firmware security, side-channel resistance, and supply-chain integrity. Moreover, TEEs and HSMs do not inherently address compromise at scale; once a device is breached, all keys stored within it may be exposed. Consequently, while HSM/TEE-based storage is a valuable defensive layer, it should be best considered as complementary rather than sufficient for PQ resilience, especially for high-value or long-lived keys.

\subsubsection{Threshold-Protected Server-Side Keys}
Threshold-based storage distributes trust by ensuring that no single server ever holds a complete cryptographic key. Instead, keys are split into shares using secret sharing or distributed key-generation techniques, and cryptographic operations require cooperation among a quorum of servers. This approach is particularly attractive for PQ settings because it limits the damage caused by partial compromise and enables graceful degradation under attack. In KDS or CA deployments, threshold protection can be applied to master keys, session-key derivation secrets, or signing keys. Even if an adversary compromises a subset of servers—or gains quantum capabilities—the system remains secure as long as the adversary does not exceed the corruption threshold. Threshold storage also enables proactive security techniques, such as periodic resharing or key refresh, which are useful mechanisms for mitigating long-term key exposure. The primary challenges lie in increased system complexity, coordination overhead, and the need for rigorous composability and side-channel analysis, particularly when threshold protocols are combined with PQC primitives.

\subsubsection{PQ-Safe Key Wrapping and Key Encryption Keys}
Key wrapping protects stored keys by encrypting them under a Key Encryption Key (KEK), allowing keys to be stored or transported securely even if storage media is compromised. In the PQ context, symmetric key-wrapping schemes—such as AES Key Wrap (AES-KW)—remain viable, provided that sufficiently large key sizes are used. Symmetric primitives with appropriate parameters are expected to remain secure in the presence of quantum adversaries, subject to Grover’s algorithm. Using AES-KW with 256-bit KEKs (or larger) enables PQ-safe protection of both symmetric session keys and PQ private keys at rest. Key wrapping is commonly used in conjunction with HSMs, KDS databases, backup systems, and inter-module key transport. However, key wrapping alone does not solve key-management risks: the security of wrapped keys ultimately depends on the protection and lifecycle management of the KEKs themselves. As a result, PQ-safe key wrapping is most effective when combined with short KEK lifetimes, periodic rewrapping, and threshold or hardware-protected KEK storage.

\subsubsection{Implications for PQ Architectures}
Together, these mechanisms shows that PQ key storage is not a binary choice between ``secure'' and ``insecure,'' but rather a spectrum of design options with distinct trust, performance, and recovery characteristics. It is important to emphasize that robust PQ systems must assume eventual compromise and therefore favor designs that limit key exposure duration, reduce single points of failure, and support recovery. Architectures that combine hardware isolation, threshold protection, and PQ-safe key wrapping offer the strongest protection against both classical and quantum-enabled adversaries, particularly when integrated into a lifecycle-aware key-management framework.

\subsection{Post-Compromise Recovery and Key Healing}
A critical challenge in the PQ era is recovery after partial or delayed compromise. It is important that systems must assume some keys generated today may be compromised in the future, necessitating robust recovery mechanisms. Lifecycle-aware architectures incorporate post-compromise security through proactive key refresh, re-derivation from uncompromised entropy sources, or threshold resharing protocols. Multi-path and MPC-based systems are particularly promising in this regard, as they enable ``key healing'' without requiring full system reset or universal re-enrollment.
\subsection{Summary of the Key Lifecycle Dimension}
This taxonomy dimension emphasizes that PQ security is inseparable from key lifecycle management. Architectures that just substitute PQ algorithms without rethinking key generation, rotation, storage, and recovery remain vulnerable to long-term and adaptive adversaries. By explicitly classifying systems according to how they manage keys across their full lifecycle, this taxonomy enables a more realistic assessment of quantum-resilient network designs and highlights where architectural innovation is still required.
\subsection{Research Gaps}
Our taxonomy shows several gaps when these lifecycle principles are applied to PQ network architectures. These gaps arise from the interaction between long-term quantum threats, distributed key-management infrastructures, and operational realities of modern networks.
\begin{itemize}
\item \emph{Gaps in Key Generation and Entropy Management.}
While key generation is typically treated as a one-time event, PQ systems must consider entropy degradation and bias accumulation over long operational lifetimes. Open questions include: (i) How to detect and recover from compromised or biased randomness sources; (ii)  How to combine distributed or multi-path entropy in a lifecycle-aware manner; (iii) How often entropy sources themselves should be refreshed or revalidated.
These issues are especially critical for symmetric-only and long-lived infrastructure keys.
\item  \emph{Gaps in Key Distribution and Rotation.}
One main issue is the limited support for frequent, large-scale rekeying. More specifically: (i) Existing key-distribution architectures are rarely evaluated under high-frequency 
rotation; (ii)  The cost of rotation in hierarchical, MPC-based, or multi-path systems has not been widely assessed; (iii) There are no guidelines  on how to coordinate rotation across heterogeneous domains and trust models. As a result, many systems implicitly rely on cryptoperiods that are too long for the PQ threat models.
\item  \emph{Gaps in Key Storage and Protection.} HSMs, TEEs, and key wrapping are commonly treated as sufficient safeguards. However:
(i) They do not address long-term exposure or delayed 
compromise; (ii) They often assume static trust in hardware vendors and firmware; (iii) They provide limited recovery guarantees once breached.
There is clearly a lack of lifecycle-aware storage designs that combine hardware isolation with proactive refresh, threshold protection, or re-wrapping strategies.
\item \emph{Gaps in Lifecycle Coordination Across Architectural Layers.} In practice, different lifecycle stages are often handled by different components (KDS, applications, HSMs, network controllers). Several components, models, and techniques are missing, namely: (i) End-to-end lifecycle orchestration mechanisms; (ii) Formal models that reason about lifecycle state across layers; (iii) Techniques to ensure consistent lifecycle enforcement in hybrid PQ/classical systems.
This fragmentation increases the risk of misconfiguration and latent vulnerabilities. The management of a highly-secure lifecycle can be expensive, and thus automation is critical.
\end{itemize}

%% file: DeploymentOperationalModel.tex
\section{Taxonomy by Deployment and Operational Model}
Beyond cryptographic foundations, key-distribution architectures, trust assumptions, and lifecycle management, PQ network designs are strongly influenced by where and how they are deployed and operated. In practice, deployment environments impose constraints on performance, availability, manageability, regulatory compliance, and upgrade cadence, all of which affect the applicability of PQ mechanisms. Moreover, operational realities—such as partial connectivity, intermittent trust anchors, legacy dependencies, and administrative boundaries—often drive which cryptographic and architectural choices are viable at any given time. To capture these factors, we classify PQ network design strategies according to deployment and operational modes, highlighting how different environments motivate distinct combinations of 
key-management architectures and trust models.

\subsection{Enterprise and Data-Center Deployments}
Enterprise networks and data centers are typically characterized by strong administrative control, reliable connectivity, and centralized identity management. These environments often deploy Kerberos-like KDSs, internal PKIs, or hybrid authentication infrastructures. PQ transition in this setting frequently begins with upgrading authentication and key-management services, followed by gradual adoption of PQ-enabled protocols for internal and external communication.
Operational challenges in enterprise environments include coexistence with legacy systems, integration with third-party services, and maintaining availability during migration. As a result, enterprises often operate mixed deployment modes, where PQ-enabled services coexist with symmetric-only or hybrid components over extended periods, thus achieving PQ migration with staged rollouts.
\subsection{Internet-Scale and Federated Deployments}
Internet-scale systems, such as cloud services, content delivery networks, and federated identity providers, operate across organizational and jurisdictional boundaries. These deployments rely heavily on standardized protocols, PKI, and interoperability, making PQ-PKI-enabled and hybrid modes particularly relevant. However, the need for backward compatibility and cross-domain trust introduces significant operational complexity. In these environments, PQ readiness is constrained not only by cryptographic support but also by coordination among independent operators, certificate lifetimes, and protocol standardization timelines. As a result, hybrid and transitional operational modes are expected to persist longer than in closed enterprise networks.

Several widely used and emerging network protocols already serve as practical entry points for PQ migration through hybrid or mixed modes. For transport security, hybrid PQ key-exchange extensions to TLS 1.3, combining classical Diffie–Hellman with post-quantum KEMs such as CRYSTALS-Kyber, have been developed and evaluated in both standards drafts and experimental implementations~\cite{Souvatzidaki}. At the infrastructure layer, PQ hybrid key exchange for IPsec/IKEv2 is being specified by the IETF as an Internet-Draft using additional PQ KEMs alongside traditional key exchanges~\cite{Kampanakis}. For domain name security, research on PQ digital signatures in DNSSEC explores how to integrate and manage larger PQ-signature schemes into the DNS security architecture~\cite{Goertzen}. These examples illustrate both the feasibility and the practical complexities of deploying hybrid PQ cryptography in real networked systems.

\subsection{IoT, Edge, and Resource-Constrained Deployments}
IoT and edge environments present some of the most challenging deployment conditions for PQ security. Devices may have limited computational resources, long deployment lifetimes, infrequent update opportunities, and minimal trust anchors. In many such settings, symmetric-only operation is the only practical option in the near term.
Key-management architectures in this mode often rely on pre-shared keys, lightweight KDSs, or serverless multi-path mechanisms. Operational constraints, such as intermittent connectivity and physical exposure, make lifecycle management, key rotation, and compromise recovery particularly difficult. These environments motivate research into minimal-assumption, low-overhead quantum-safe designs.
\subsection{Industrial, Critical Infrastructure, and Regulated Environments}
Industrial control systems, energy grids, transportation networks, and other critical infrastructures are characterized by strict availability requirements, long certification cycles, and regulatory oversight. In these environments, cryptographic upgrades must be carefully staged, and operational risk often outweighs cryptographic novelty. As a result, deployments frequently operate in conservative or partial PQ modes, combining symmetric-only mechanisms with carefully validated PQ components. Hierarchical or threshold-based KDS architectures are attractive here, as they allow gradual hardening of trust anchors without disrupting real-time operations. Operational modes in these environments emphasize stability, auditability, and controlled evolution rather than rapid adoption.

\subsection{Mobile, Wireless, and Next-Generation Network Deployments}
Mobile and wireless networks, including 5G, emerging 6G concepts, and next-generation access networks, operate under stringent latency, scalability, and mobility constraints. Authentication and key establishment are often tightly integrated into network control planes, making architectural changes particularly impactful.
In these deployments, PQ transition may involve hybrid operational modes, where symmetric-only core mechanisms are augmented with PQ-enabled authentication at selected interfaces. Operational activities such as handover frequency, roaming agreements, and multi-operator trust relationships further complicate deployment. These environments highlight the importance of aligning PQ mechanisms with performance and availability requirements.

\subsection{Disconnected, Intermittent, and Adversarial Deployments}
Some deployments, such as tactical networks, emergency response systems, or remote sensing platforms, must operate with limited connectivity, minimal infrastructure, and strong adversarial assumptions. In these settings, reliance on continuously available trusted servers is impractical. Serverless multi-path key distribution, pre-distributed symmetric trust, and opportunistic rekeying are particularly relevant operational modes. These deployments often combine symmetric only cryptography with trust-minimized or zero-trust assumptions, emphasizing resilience, autonomy, and post-compromise recovery over centralized control.

\subsection{Summary of Deployment and Operational Dimension}
\input{table7}
This taxonomy dimension underscores that PQ readiness is strictly  dependent on deployment context and operational constraints. 
Table 2 summarizes which key-distribution architectures are most suitable across common deployment environments, highlighting how operational constraints and threat models shape feasible post-quantum design choices.
As we can see from the table, different environments require different cryptographic foundations, trust models, and key-management architectures, and most real-world systems will operate across multiple deployment modes simultaneously. Recognizing and explicitly modeling these operational differences is essential for designing transition strategies that are both secure and practical in heterogeneous post-quantum landscapes.

\begin{center}
\begin{tcolorbox}[
  enhanced,
  width=0.96\linewidth,
  colback=gray!10,
  colframe=black,
  arc=4pt,
  boxrule=0.8pt,
  left=6pt,
  right=6pt,
  top=6pt,
  bottom=6pt
]
\textbf{Operational Reality: PQ Transition Is Inherently Heterogeneous}

Real-world networks rarely transition to PQ security uniformly. Enterprises, cloud providers, IoT deployments, and critical infrastructure often operate simultaneously under symmetric-only, hybrid, and PQ-enabled modes. Safe operation during these prolonged intermediate states—rather than the final ``fully PQ'' endpoint—represents the dominant operational challenge in post-quantum migration.

\end{tcolorbox}
\end{center}

\subsection{Research Gaps}
While the taxonomy by deployment and operational model captures how PQ mechanisms are realized in practice across enterprises, cloud infrastructures, IoT, critical systems, and mobile networks, it also reveals significant gaps between theoretical PQ security designs and their operational viability. These gaps originated from the reality that deployment environments differ widely in update frequency, trust boundaries, performance constraints, and tolerance for disruption—factors that are often underrepresented in cryptographic and protocol-centric research.
\\
\textbf{Gaps in Managing Heterogeneous PQ Readiness}
\begin{itemize}
\item \emph{Lack of formal models for mixed-capability deployments.} Most PQ designs implicitly assume homogeneous environments in which all components can adopt new cryptographic mechanisms simultaneously. In practice, deployments are heterogeneous, with PQ-capable systems coexisting alongside legacy and constrained components for extended periods. There is a lack of formal models that characterize: (i) security guarantees under partial PQ adoption; (ii) risks introduced by long-term coexistence of cryptographic modes; (iii)
conditions under which heterogeneity degrades overall system security.
\item \emph{Insufficient guidance for safe coexistence.} 
Current transition guidance focuses on migration endpoints rather than the prolonged intermediate states that dominate real deployments. How to safely operate hybrid, symmetric-only, and PQ-enabled components concurrently—without introducing downgrade or lateral-movement vulnerabilities—remains poorly understood.

\end{itemize}
\textbf{Gaps in Operational Lifecycle Integration}
\begin{itemize}
\item \emph{Weak coupling between deployment models and key lifecycle management.} Although key lifecycle management is widely recognized as critical, its interaction with deployment models is rarely analyzed. For example: (i) frequent rekeying may be feasible in cloud environments but impractical in industrial or IoT deployments; (ii)
post-compromise recovery strategies vary significantly across operational contexts; (iii)
lifecycle guarantees often break down at organizational or administrative boundaries.
There is a lack of deployment-aware lifecycle frameworks that adapt key management practices to operational constraints.

\item \emph{Limited treatment of long-lived systems.} Many operational environments, such as industrial control systems, embedded devices, and infrastructure networks, are designed for decades-long lifetimes. Existing PQ transition strategies rarely address how such systems can evolve cryptographically over time without violating safety, certification, or availability requirements.
\end{itemize}

\textbf{Gaps in Availability and Reliability Under PQ Transition}
\begin{itemize}
    \item \emph{Under-investigated impact of PQ mechanisms on operational stability.} PQ mechanisms often introduce increased computational overhead, larger messages, and more complex handshakes. There is limited empirical understanding of how these changes affect: (i) availability and latency in real deployments; (ii) failure modes under load or attack; (iii)  operational stability during rolling upgrades or partial failures.
    \item \emph{Single points of operational fragility.} Many deployment models rely on centralized services, such as key-management servers, identity providers, or trust anchors. While these components are natural upgrade targets, their compromise or unavailability can have a major impact. Research on deployment models that balance centralization with resilience, particularly under PQ threat assumptions, remains limited.
\end{itemize}

\textbf{Gaps in Cross-Domain and Federated Operations}
\begin{itemize}
    \item \emph{Lack of deployment-aware trust federation models.} Current network systems frequently span organizational, regulatory, and jurisdictional boundaries. PQ deployment strategies must therefore address trust federation, inter-domain authentication, and policy coordination. However, existing work rarely considers how differing PQ readiness levels across domains affect overall security or how trust should be renegotiated during transition.
    \item \emph{Insufficient analysis of operational governance.} Deployment and operational models are shaped not only by technology but also by governance, policy, and compliance. There is little research on how cryptographic agility, key lifecycle enforcement, and PQ guarantees can be governed consistently across heterogeneous operational environments.
\end{itemize}

\textbf{Gaps in Deployment-Aware Cryptographic Agility}
\begin{itemize}
\item \emph{Emphasis on software-level agility only.} Most discussions of cryptographic agility focus on algorithm substitution within software stacks. This perspective overlooks the operational realities of networks, where agility depends on: (i)
architectural flexibility; (ii) key-management infrastructure; (iii) protocol layering and interoperability; (iv) tolerance for mixed cryptographic states.
There is a lack of frameworks that define cryptographic agility as a deployment-level and operational property.

\item \emph{Absence of metrics for operational agility.} Finally, there are no widely accepted metrics for evaluating how agile a deployment model truly is. Questions such as ``How quickly can this network rotate trust anchors?'' or ``How many components must be upgraded to eliminate a legacy algorithm?'' remain largely qualitative.
\end{itemize}

Overall, this taxonomy shows that PQ security is constrained as much by deployment and operational realities as by cryptographic design. Addressing these gaps requires research that bridges cryptography, systems engineering, and operations, with explicit attention to heterogeneity, lifecycle integration, availability, and governance. Without such work, PQ mechanisms risk remaining theoretically sound but operationally brittle in real-world deployments.

%% file: table7.tex
\begin{table*}[t]
\centering
\footnotesize
\renewcommand{\arraystretch}{0.95}
\caption{Deployment Environments and Suitable Post-Quantum--Resistant Architectures}
\label{tab:deployment-architectures}
\begin{tabular}{|p{3.2cm}|p{3.4cm}|p{4.0cm}|p{3.4cm}|}
\hline
\textbf{Deployment Environment} &
\textbf{Dominant Constraints} &
\textbf{Suitable Architectures} &
\textbf{Notes} \\
\hline

Enterprise / Data Center &
Central administrative control; legacy integration &
Centralized KDS, Hierarchical KDS, Threshold/MPC-based KDS &
Hybrid PQ rollout common; gradual lifecycle hardening \\
\hline

Internet-scale / Federated &
Interoperability; PKI dependence; cross-domain trust &
PQ-PKI-based, Hybrid PQ, Hierarchical KDS &
Long-term coexistence of classical and PQ mechanisms \\
\hline

IoT / Edge &
Resource constraints; long device lifetimes &
Symmetric-only KDS, Serverless multi-path &
PQ-PKC often infeasible; key lifecycle critical \\
\hline

Industrial / Critical Infrastructure &
Availability; certification; safety requirements &
Hierarchical KDS, Threshold/MPC-based KDS &
Conservative migration; strong emphasis on recovery \\
\hline

Mobile / Next-Generation Networks &
Mobility; latency; handover frequency &
Contributory GKE, Hierarchical KDS &
Path diversity and trust vary across operators \\
\hline

Adversarial / Zero-Trust Environments &
Strong network attackers; minimal trust assumptions &
Threshold/MPC-based KDS, Serverless multi-path &
Trust minimization and compromise containment key \\
\hline

\end{tabular}
\end{table*}

%% file: CommunicationTopologyNetworkLayer.tex
\section{Taxonomy by Network Communication Topology and Network  Layer}
PQ security mechanisms do not operate in a vacuum: their feasibility, performance, and security guarantees are shaped jointly by the network layer at which they are deployed and by the communication topology over which they operate. While protocol layering determines where cryptographic protections are enforced (e.g., link, network, transport, or application layer), communication topology determines how traffic flows, which intermediaries are involved, and what adversarial capabilities are realistic. In practice, these two aspects are deeply intertwined and must be considered together.

In the discussion in this section we 
therefore consider communication topology as a cross-cutting factor that impacts the security and deployability of PQ mechanisms at every layer of the network stack. Topological characteristics—such as single-hop versus multi-hop paths, intra-domain versus inter-domain routing, path stability, mobility, and administrative boundaries—directly affect assumptions about trust, path independence, key lifetimes, and the availability of key-management services. For example, serverless multi-path key-distribution schemes rely on path diversity assumptions that may hold in wide-area, multi-AS settings but break down in centralized cloud backbones. Similarly, highly dynamic or mobile topologies constrain handshake complexity and limit reliance on continuously available trusted servers.

Within this context, network layering provides the organizational structure for analyzing PQ mechanisms. Each layer introduces distinct constraints on latency, state management, hardware acceleration, and upgrade cadence, while inheriting and amplifying the effects of the underlying topology. Lower layers tend to operate within stable, tightly controlled topologies, favoring symmetric cryptography and hop-by-hop protection. Higher layers abstract away routing details but must tolerate heterogeneous paths, federated trust, and partial upgrade adoption across endpoints.

The remainder of this section therefore proceeds layer by layer, from link-layer mechanisms to application-level protocols, while explicitly accounting for how topological realities shape post-quantum design choices at each layer. Rather than treating topology as an independent dimension, we show how it conditions the effectiveness of PQ approaches across the stack—highlighting where certain assumptions are reasonable, where they are fragile, and where new protocol designs are required to bridge the gap between cryptographic theory and networked reality. More in details:
\begin{itemize}
\item At Layer 2, we focus on environments characterized by stable, single-domain topologies, where hop-by-hop protection and hardware acceleration dominate design considerations.
\item
At Layer 3, we move to multi-hop and inter-domain topologies, where routing dynamics, AS-level adversaries, and path correlation become central to Q security analysis.
\item
At Layer 4, we consider end-to-end transport protocols operating over arbitrary and often heterogeneous paths, emphasizing session establishment, mobility, and performance trade-offs.
\item
Finally, at the application layer, we analyze protocols that define their own logical topologies—often spanning administrative domains—and must reconcile PQ security with usability, federation, and long-lived trust relationships.
\end{itemize}

By structuring the discussion in this way, we make explicit how communication topology conditions the deployment and security of post-quantum mechanisms at every network layer, and why effective post-quantum transition strategies must reason jointly about topology, layering, and protocol evolution rather than addressing them in isolation.

\subsection{Network Layers}
PQ transition is not uniform across the network stack. Different layers impose distinct constraints on latency, state management, hardware acceleration, trust boundaries, and upgrade cadence. As a result, PQ–resistant mechanisms must be evaluated layer by layer, taking into account both cryptographic feasibility and operational impact. This taxonomy dimension classifies PQ approaches according to the network layer at which cryptographic protection is applied, while explicitly accounting for communication topology and emerging protocol designs.

\subsubsection{Layer 2: Switching and MAC Layer}
At the link layer, cryptographic mechanisms operate over stable, hop-by-hop communication topologies, typically within a single administrative domain, where low latency, hardware acceleration, and tightly controlled trust assumptions are critical factors in  design choices.
Possible PQ approaches include:
\begin{itemize}
\item \emph{PQ-Secure MACsec.} MACsec (IEEE 802.1AE) currently relies on symmetric cryptography (e.g., AES-GCM) for frame protection. From a PQ perspective, symmetric primitives with sufficiently large keys are expected to remain secure against quantum adversaries. As a result, MACsec can be considered quantum-resistant in its data-plane cryptography, provided that appropriate key sizes are used.
The main PQ challenge at this layer is not in encryption itself, but in key establishment and rekeying. MACsec deployments depend on key agreement protocols (e.g., 802.1X) that may involve classical PKC. Transitioning these control-plane components to PQ-safe or symmetric-only alternatives remains an open problem, particularly in large-scale switching fabrics.
\item \emph{Link-Layer Hop-by-Hop Keying.}
In tightly controlled environments, link-layer keys may be pre-provisioned or derived via centralized key-management systems. While this approach avoids PQC entirely, it raises scalability and lifecycle concerns. Frequent rekeying is essential to limit exposure, but must be carefully engineered to avoid disrupting forwarding performance.
\end{itemize}

\subsubsection{Layer 3: IP Layer}
At the network layer, PQ security mechanisms must operate over multi-hop, routed topologies that often span multiple administrative domains, making routing dynamics, path correlation, and AS-level adversaries central to the security analysis and design choices.
Possible PQ approaches include:
\begin{itemize}
\item \emph{PQ-Enabled IPsec.}
IPsec is an obvious candidate for PQ transition, as it already separates key establishment (IKE) from data-plane protection. While the data plane relies on symmetric cryptography, IKE traditionally depends on Diffie–Hellman and signatures, both of which must be replaced or augmented in a post-quantum setting. PQ-enabled or hybrid IKE variants are therefore central to IP-layer post-quantum security.
Operational challenges include increased handshake size, potential latency overhead, and interoperability across heterogeneous endpoints. Moreover, IPsec deployments often involve long-lived tunnels, which heightens the importance of robust key rotation and rekeying policies under PQ threat models.
\item \emph{Multi-Path IP Routing for Secret Sharing.} At the IP layer, multi-path routing enables serverless or trust-minimized key distribution by splitting key material across multiple routes. This approach can provide strong protection against adversaries who do not control all paths simultaneously. However, its effectiveness depends on assumptions about routing diversity and independence across ASs. Real-world routing dynamics, traffic engineering, and centralized cloud backbones complicate these assumptions.
\end{itemize}

\subsection{Layer 4: Transport Layer}
The transport layer is a focal point of PQ transition because it naturally encapsulates session semantics and endpoint authentication.
At the transport layer, cryptographic protections are applied end-to-end across heterogeneous and potentially changing network paths, abstracting away routing details while inheriting the topological complexity of the underlying network. We discuss several PQ approaches below.

\emph {PQ TLS.} TLS is the most widely deployed secure transport protocol, and its PQ evolution has been extensively studied. PQ-TLS variants replace or hybridize classical key exchange with post-quantum KEMs, while continuing to rely on symmetric cryptography for bulk data protection. The transport layer provides a convenient abstraction for managing cryptographic agility, certificate lifetimes, and session resumption. Key challenges include increased handshake overhead, certificate size inflation, and the need to coordinate transition across diverse client populations. Moreover, TLS often serves as a dependency for higher-layer protocols, amplifying the impact of design choices made at this layer.

\emph{PQ QUIC.} QUIC integrates transport and security more tightly than TLS-over-TCP, offering improved performance and resilience. PQ-enabled QUIC inherits many challenges from PQ-TLS but must additionally account for connection migration, stateless retries, and 0-RTT data. These features complicate the design of PQ-safe handshakes and raise subtle security questions regarding replay and downgrade resistance.

\subsection{Application-Layer Protocols}
At the application layer, security protocols define their own logical communication topologies, often spanning federated administrative domains and long-lived trust relationships that are only loosely coupled to the underlying network paths.
Application-layer protocols often embed security mechanisms tailored to specific use cases, making PQ transition highly context-dependent. Possible PQ approaches are discussed below.

 \emph{PQ SSH.} SSH relies on public-key cryptography for both authentication and key exchange, making it directly impacted by quantum threats. PQ-enabled SSH variants explore replacing classical key exchange with PQ KEMs, while maintaining backward compatibility through hybrid modes. Because SSH is widely used for administrative access, transitioning this protocol is operationally sensitive and must balance security improvements against deployment risk.
 
\emph{Secure Messaging and End-to-End Protocols.} 
End-to-end encrypted messaging protocols (e.g., Signal-like designs) introduce additional constraints, such as asynchronous key establishment and long-term identity keys. PQ adaptations must carefully manage key size, statefulness, and forward secrecy, while preserving usability and bandwidth efficiency.

Apple’s PQ3 protocol integrates post-quantum cryptography not only into the initial key exchange but also into continuous key evolution, aiming to provide forward secrecy and post-compromise security even against adversaries with future quantum capabilities. A formal analysis by David Basin et al.~\cite{basin_iMessage_usenix2025} models PQ3’s multi-stage key updates and shows that its design achieves strong security under active attacker models, including robustness of its ratcheting mechanism when parties are compromised and later recover. 

Signal’s post-quantum transition begins with PQXDH, a hybrid key exchange that combines classical elliptic-curve Diffie–Hellman with a post-quantum KEM to protect initial session establishment against “harvest now, decrypt later” attacks. A formal, machine-checked analysis by Karthikeyan Bhargavan et al. \cite{pqxdh_usenix2024} shows that PQXDH achieves strong authentication and secrecy guarantees in a symbolic model, while preserving compatibility with the existing Signal protocol design. More recent work (SPQR) \cite{signal_spqr_2025} extends these guarantees beyond the initial handshake by introducing post-quantum ratcheting, though this is still an evolving deployment.

\emph{Application-Level VPNs and V2X Keying.} Application-layer VPNs and vehicle-to-everything (V2X) communication systems often operate in highly dynamic and safety-critical environments. PQ transition in these contexts must accommodate low latency, intermittent connectivity, and regulatory requirements. Hybrid and symmetric-heavy designs are therefore likely to persist, with PQ mechanisms introduced incrementally as standards mature.






\subsection{Summary of by Taxonomy Network 
Layer}
The dimension on protocol layers shows that PQ transition is inherently layer-specific. Lower layers favor symmetric cryptography and stable trust relationships but face challenges in key management and hardware integration. Transport and application layers offer greater flexibility for PQ adoption but introduce performance and interoperability concerns. Emerging protocols further complicate the landscape by tightly coupling security with transport semantics and application logic. A realistic PQ strategy must therefore coordinate mechanisms across layers, rather than treating protocol upgrades in isolation.

\subsection{Research Gaps}
While the taxonomy clarifies how PQ mechanisms interact with the different network layers, it also shows several challenges that are not addressed in existing literature. These gaps originates from the fact that most PQ research considers cryptographic protocols largely in isolation, whereas real networks are shaped by routing dynamics, mobility, hardware constraints, and heterogeneous administrative control.

\textbf{Gaps in Topology-Aware Threat Modeling}
\begin{itemize}
\item \emph{Lack of integrated cryptographic–network adversary models.} Most PQ analyses focus on cryptographic adversaries and abstract away the network. However, networking security models often assume classical cryptography. There is a lack of unified adversary models able to simultaneously capture: (i) quantum-capable cryptanalysis; (ii)  AS-level traffic observation or manipulation; (iii) selective path compromise or correlation; (iv) long-term traffic recording (``harvest-now, decrypt-later''). 
Without such models, it remains unclear which PQ mechanisms remain secure under realistic topological conditions.
\item \emph{Unvalidated assumptions about path diversity.} Multi-path and serverless key-distribution approaches rely on assumptions of path independence that may not hold in modern networks dominated by cloud backbones, traffic engineering, and centralized control. There is little empirical work quantifying effective path diversity across real-world topologies or measuring how many independent paths are required to achieve meaningful security gains.
\end{itemize}

\textbf{Gaps at Lower Network Layers (L2/L3)}
\begin{itemize}
\item \emph{Underexplored PQ transition at the link layer.}
While link-layer encryption relies on symmetric cryptography and is often considered quantum-resistant, little attention has been paid to the PQ security of link-layer key management, including: (i)
secure bootstrapping of link keys; (ii) rekeying under high churn; (iii) interactions with centralized authentication. These issues are particularly relevant for large switching fabrics and data-center networks. 

\item \emph{Limited understanding of PQ-enabled IP-layer security at scale.} PQ-enabled IPsec and related mechanisms face challenges in handshake size, rekey frequency, and interoperability across heterogeneous endpoints. There is a lack of large-scale evaluations examining how PQ mechanisms affect: (i) tunnel establishment latency; (ii)  routing stability; (iii) operational complexity in multi-domain deployments.
\end{itemize}

\textbf{Gaps at the Transport Layer}
\begin{itemize}
\item \emph{Incomplete analysis of PQ handshake impact under realistic topologies.}
Transport-layer protocols abstract routing, but still inherit the performance and reliability characteristics of the underlying network. Existing work on PQ-TLS and PQ-QUIC often evaluates isolated handshakes, leaving open issues concerning: (i) performance under lossy or mobile paths; (ii) interaction with middleboxes and load balancers; (iii) scalability in environments with frequent short-lived connections.

\item \emph{Limited support for topology-induced heterogeneity.} Transport protocols increasingly operate (will operate) in environments where some endpoints are PQ-capable while others are not. There is no framework for reasoning about mixed-capability deployments, downgrade resistance, or security guarantees during prolonged coexistence of cryptographic modes.
\end{itemize}

\textbf{Gaps at the Application Layer and Emerging Protocols}
\begin{itemize}

\item \emph{Fragmented PQ transition across application protocols.} 
Application-layer protocols such as SSH, secure messaging, VPNs, and V2X systems embed cryptography in protocol-specific ways. As a result: (i) PQ transition strategies are inconsistent across applications; (ii) lifecycle and trust assumptions vary widely; (iii) security guarantees become difficult to compare or compose.
There is no unifying framework for coordinating PQ transition across application protocols operating over a shared infrastructure.

\item \emph{Lack of impact analysis of mobility and federation.} Next-generation protocols increasingly support mobility, federation, and cross-organizational trust. How PQ mechanisms interact with roaming, identity federation, and dynamic trust establishment remains largely unexplored, particularly under adversarial network conditions.
\end{itemize}

\textbf{Cross-Layer and Cross-Topology Gaps}

\begin{itemize}
\item \emph{Absence of end-to-end reasoning across layers.}  PQ mechanisms are typically analyzed at a single layer. However real security emerges from their composition across layers. There is a lack of formal and empirical methods for reasoning about end-to-end security when different layers adopt different cryptographic foundations, key lifecycles, and trust models.

\item \emph{Lack of topology-aware transition strategies.}
Current transition guidance focuses on protocol upgrades rather than on how those upgrades behave under different topologies. For example, strategies that work well in stable enterprise networks may fail in mobile, multi-domain, or adversarial environments. Explicitly topology-aware transition frameworks are missing.
\end{itemize}

Overall, this analysis shows that PQ readiness at the network level is constrained as much by topology and layering as by cryptographic choice. Addressing these gaps will require research that integrates cryptography, network measurement, protocol design, and systems engineering. Without such integration, PQ mechanisms risk being secure in theory but fragile in practice.

%% file: SecurityProperties.tex
\section{Taxonomy by Security Properties}
In addition to architectural and deployment considerations, PQ networks must be evaluated according to the security properties they provide under realistic threat models. In the PQ era, traditional properties such as confidentiality and authentication must be reconsidered in the presence of long-lived adversaries, delayed compromise detection, and powerful network-level attackers. This taxonomy classifies PQ network designs according to a set of security properties that are particularly relevant for distributed, heterogeneous, and long-lived systems.

\subsection{Perfect Forward Secrecy}
Perfect Forward Secrecy (PFS) ensures that compromise of long-term keys does not expose past session traffic. This property is especially critical in the PQ setting, where adversaries may record encrypted traffic for extended periods and attempt decryption once cryptanalytic capabilities improve.
In static symmetric key-distribution systems, such as traditional Kerberos deployments, PFS is inherently difficult to achieve. Long-term shared secrets and ticket-granting keys create dependencies across sessions, so compromise of a central authority or a client key can retroactively expose large volumes of traffic.
By contrast, KEM-based key-establishment protocols naturally support PFS by generating ephemeral session keys that are cryptographically independent of long-term credentials. Hybrid and PQ-enabled transport protocols therefore offer stronger forward secrecy guarantees, provided that ephemeral keys are properly managed and rotated.

In symmetric-only environments, PFS can be approximated using multi-path key establishment combined with hash-chain–based key evolution. By deriving session keys from fresh entropy aggregated across multiple paths and advancing keys via one-way functions, systems can limit retrospective exposure even without public-key cryptography.

\subsection{Post-Compromise Security and Healing}
Post-compromise security extends beyond forward secrecy by addressing the system’s ability to recover after a partial or delayed compromise. In large-scale networks, compromise detection may occur long after the initial breach, making full system reset impractical.

Centralized key-distribution systems offer limited recovery options: once a master key or KDS is compromised, all dependent keys must be reissued, often requiring disruptive re-enrollment. This motivates architectures that limit the impact of compromise and enable incremental recovery.
Threshold and MPC-based KDS architectures improve post-compromise security by ensuring that no single server compromise reveals complete secrets. Proactive resharing and distributed rekeying allow systems to ``recover'' over time without global reset. Similarly, multi-path secret sharing enables recovery by injecting fresh entropy from uncompromised paths. Hash-chain–based rapid rekeying mechanisms further support post-compromise recovery by ensuring that compromised keys cannot be used to derive future keys, even if compromise persists for some period.

\subsection{Key Compromise Impersonation Resistance}
Key Compromise Impersonation (KCI) resistance captures whether an adversary who compromises a party’s long-term key can subsequently impersonate other parties to the victim. This property is particularly relevant in PQ settings where key compromise may be silent and long-lasting.
In symmetric-only systems, KCI resistance is inherently weak: possession of a shared secret often enables impersonation in multiple directions. Static symmetric credentials therefore amplify the impact of compromise.
KEM-based protocols offer stronger KCI resistance by separating long-term authentication credentials from ephemeral session keys, limiting the attacker’s ability to impersonate peers. However, improper protocol composition or reuse of keys across roles can still undermine this property.
Architectural techniques can further mitigate KCI risk. MPC-based KDSs prevent attackers from learning full impersonation credentials, while multi-path key establishment reduces correlated exposure by requiring adversaries to compromise multiple independent components to mount successful impersonation attacks.

\subsection{Resistance to Quantum Side-Channel Attacks}
Even quantum-resistant cryptographic algorithms may be vulnerable to side-channel attacks, including timing, power, cache, or microarchitectural leakage. In the PQ era, this risk is amplified by the increased complexity of many PQ primitives.
PQ public-key implementations require careful software and hardware hardening to prevent leakage of secret material during key generation and decapsulation. These requirements pose challenges for constrained devices and latency-sensitive environments.

In contrast, symmetric-only solutions often rely on simpler, well-understood primitives that are easier to harden and benefit from decades of implementation experience. This makes symmetric-heavy designs attractive in environments where side-channel resistance and implementation simplicity are paramount, even if other security properties must be carefully engineered.

\subsection{Network-Level Adversary Resistance}
Many PQ analyses assume purely cryptographic adversaries, but real networks face active network-level attackers capable of man-in-the-middle attacks, route manipulation, traffic correlation, and selective denial of service.
Protocols that rely on centralized trust anchors or single communication paths are particularly vulnerable under such adversarial models. For example, routing manipulation can undermine assumptions about path independence or force traffic through compromised intermediaries.
Multi-path communication and key establishment offer a promising defense by exploiting path diversity to reduce adversarial visibility and control. However, their effectiveness depends critically on the underlying topology and the attacker’s ability to influence routing decisions, especially at the AS level.
Designing PQ protocols that explicitly account for network-level adversaries, rather than treating the network as a transparent channel, remains a key open challenge.

\subsection{Summary of Security Property Dimension}
This taxonomy highlights that PQ security is not a single property but a set of interrelated guarantees. PFS, post-compromise recovery, KCI resistance, side-channel resilience, and resistance to network-level adversaries are achieved through different combinations of cryptographic primitives, architectures, and deployment choices. Evaluating post-quantum network designs along these dimensions enables more nuanced comparison and exposes trade-offs that are invisible when focusing on algorithmic quantum resistance alone.

\subsection{Research Gaps}
While the taxonomy by security properties clarifies the goals that PQ networks should aim to achieve, it also exposes significant gaps between desired security guarantees and what current protocols and architectures can realistically provide. These gaps arise from fundamental tensions between cryptographic strength, architectural complexity, and operational constraints in heterogeneous networks.

\textbf{Gaps in PFS}
\begin{itemize}
\item \emph{Lack of PFS in symmetric-only, centralized systems.} Traditional symmetric key-distribution systems, such as static Kerberos-like architectures, cannot provide strong PFS without significant architectural modification. Although techniques such as hash-chain–based rekeying and multi-path entropy injection can approximate PFS, there is no widely accepted framework that formalizes the security guarantees these constructions provide under PQ adversaries.
\item emph{Limited analysis of PFS under network compromise.} Existing PFS analyses typically assume honest-but-curious networks. There is little work analyzing whether PFS guarantees hold when adversaries can manipulate routing, delay messages, or correlate multi-path traffic—conditions that are realistic in wide-area and inter-domain networks.
\end{itemize}

\textbf{Gaps in Post-Compromise Security and Key Healing}
\begin{itemize}
\item \emph{Absence of standardized post-compromise recovery models.} While threshold KDSs, MPC-based designs, and multi-path key establishment suggest promising approaches to post-compromise recovery, there is no standardized model for how recovery should proceed in practice. Open questions include: (i) How quickly systems can regain security after compromise; (ii)  What guarantees hold during the recovery window; (iii) How recovery interacts with long-lived sessions and cached credentials.
\item \emph{Limited support for recovery at scale.}
Most post-compromise security mechanisms are evaluated in small or idealized settings. There is little empirical evidence demonstrating that key healing techniques remain effective in large-scale, multi-domain networks with millions of endpoints and diverse trust assumptions.
\end{itemize}

\textbf{Gaps in KCI Resistance}
\begin{itemize}
\item \emph{Insufficient KCI analysis for symmetric-only architectures.} KCI resistance has been extensively investigated for public-key protocols but remains poorly understood in symmetric-only settings. In particular, there is a lack of formal analysis quantifying how architectural choices—such as centralized vs. threshold KDSs—affect impersonation risk after key compromise.
\item \emph{Incomplete treatment of KCI in hybrid and multi-path systems.} Hybrid PQ and multi-path designs combine multiple cryptographic mechanisms, yet their KCI properties are rarely analyzed compositionally. It remains unclear whether combining mechanisms improves KCI resistance or introduces subtle vulnerabilities due to protocol interaction effects.
\end{itemize}

\textbf{Gaps in Resistance to Quantum Side-Channel Attacks}
\begin{itemize}
\item \emph{ Limited real-world evaluation of PQ side-channel resilience.}
Although side-channel attacks are widely recognized as a major risk for PQ implementations, most evaluations remain theoretical or laboratory-based. There is a lack of systematic studies assessing side-channel leakage in real-world deployments, particularly in resource-constrained devices and shared cloud environments.

\item \emph{Lack of lifecycle-aware side-channel mitigation.} Current defenses focus on hardening individual implementations, but do not address how side-channel risks evolve over time. For example, key rotation, threshold resharing, and migration across cryptographic primitives may inadvertently introduce new leakage vectors that are not captured by static analysis.
\end{itemize}

\textbf{Gaps in Resistance to Network-Level Adversaries}
\begin{itemize}
\item \emph{Absence of unified models for cryptographic and network attacks.}
Most PQ protocols are analyzed under cryptographic adversary models that abstract away the network. Conversely, network-security models rarely incorporate quantum capabilities. There is a lack of unified frameworks that combine: (i) quantum-capable cryptanalysis; (ii) active man-in-the-middle attacks; (iii) route manipulation and path correlation; (iv)  long-term traffic observation.

\item \emph{Fragility of multi-path defenses under realistic routing conditions.}
Multi-path approaches promise strong security against network adversaries, but their effectiveness depends on assumptions that may not hold in practice. There is little empirical work validating whether sufficient path diversity exists in modern networks, particularly in environments dominated by large cloud providers or centralized routing policies.
\end{itemize}

Overall, this taxonomy reveals that many security properties crucial for PQ networks, such as PFS, post-compromise recovery, KCI resistance, and robustness against network-level adversaries, remain incompletely understood or insufficiently realized in practice. Addressing these gaps will require not only new cryptographic constructions, but also architectural innovation, realistic adversary modeling, and large-scale empirical validation. Without such advances, PQ security risks remaining robust in theory but brittle in real-world deployments.

%% file: BestPractices.tex
\section{Best Practices}

The PQ transition is not a single cryptographic event but a prolonged architectural transformation. Best practices for quantum-resistant networks must, therefore, extend beyond algorithm substitution and address trust models, key-distribution architectures, lifecycle management, and deployment realities. The following practices synthesize guidance from NIST post-quantum transition documents (e.g., NIST SP 800-208, SP 800-56C Rev. 2, SP 800-227), IEEE security standards, and emerging industry migration roadmaps, interpreted through the architectural taxonomy presented in this article. The goal is not to prescribe a single solution, but rather to provide \emph{context-sensitive guidance} that aligns cryptographic choices with operational constraints. 

\subsection{Cryptographic Discovery and Migration Roadmap}
If the organization does not have a complete and continuously maintained view of where cryptography is used (for example, in protocol endpoints, device identities, key establishment, signatures, embedded firmware validation, VPNs, etc.), as a best practice, it is actually necessary to carry out cryptographic discovery, since this is the prerequisite for every other organizational transition decision. In practice, this means building an inventory that is specific enough to answer questions such as what algorithms are used, for what purpose, in what products/services/platforms, and with what data lifetime assumptions. This documentation becomes the basis for prioritization and for avoiding migration surprises. NCCoE’s\footnote{https://www.nccoe.nist.gov/crypto-agility-considerations-migrating-post-quantum-cryptographic-algorithms} migration workstream explicitly foregrounds cryptographic inventory/discovery as a core enabling capability for prioritizing where to implement PQC first. When there is uncertainty about ``how urgent'' PQC transition is, the operational stance recommended by NIST's migration roadmap is to plan now, especially for high-value and long-lived data and for systems with long replacement cycles. 

\subsection{Design for Cryptographic Agility at the Architectural Level}
For any system that is expected to live through multiple cryptographic transitions (which is the realistic case for networks and infrastructure), crypto agility is a first-class requirement rather than a future enhancement. Indeed, NIST and IEEE guidance consistently emphasize cryptographic agility as a prerequisite for PQ readiness, but often frame it narrowly as the ability to replace algorithms within protocols. Our analysis shows that this view is insufficient for networks. In practice, cryptographic agility must be treated as an architectural property, encompassing the ability to evolve trust anchors, key-distribution mechanisms, and lifecycle policies without service disruption. NIST CSWP 39\footnote{https://nvlpubs.nist.gov/nistpubs/CSWP/NIST.CSWP.39.pdf}, published recently, frames crypto agility as the capability to replace and adapt cryptography across protocols, applications, hardware, and infrastructures without unacceptable disruption, and it highlights that agility requirements differ across implementation environments (e.g., embedded vs cloud-native). 
Thus, when designing or selecting protocols that must support PQC, it is important to avoid agility designs that create combinatorial complexity or negotiation ambiguity. CSWP 39 discusses the tradeoff between (a) cipher-suite style identifiers and (b) multiple independent algorithm identifiers; either can work, but mixing approaches inconsistently or leaving compatibility logic underspecified tends to produce operational fragility. Therefore, it is necessary to select a consistent identification/negotiation strategy and specify acceptable combinations explicitly.

Cryptographic agility in PQ networks should be treated as a design objective, not simply as a protocol feature. Organizations should distinguish between algorithm agility (the ability to replace cryptographic primitives) and architectural agility (the ability to evolve trust models, key-distribution mechanisms, and lifecycle policies without service disruption.) In practice, architectural agility requires: (i) support for hybrid and symmetric-only fallback modes; (ii) short cryptoperiods and automated key rotation; (iii) mechanisms for post-compromise recovery and re-enrollment; and (iv) tolerance for heterogeneous cryptographic capabilities across devices and domains. Networks that lack these properties may remain brittle even after adopting PQ algorithms. As a result, transition planning should prioritize architectures that enable staged migration and coexistence, rather than assuming a single, global cryptographic upgrade event.

In general, from the agility perspective, best practice would be to decouple ``what the system does'' from ``which algorithms it uses.'' Where possible, standardize internal APIs, configuration surfaces, and deployment processes so that algorithm changes are policy-driven rather than code-driven. This aligns directly with CSWP 39’s treatment of crypto agility in system implementations (e.g., library APIs, OS/kernel contexts, cloud-native/service mesh patterns, embedded constraints, and hardware realities). Furthermore, design networks that can support hybrid cryptography, symmetric-only fallback modes, and gradual trust evolution is critical. This aligns with NIST's emphasis on avoiding long-lived dependencies on any single cryptographic assumption, while extending it to networked systems where keys, servers, and topology interact. Architectures that presume a single trust model, such as a monolithic PKI or a static centralized KDS, may support algorithmic agility in theory but remain operationally brittle during PQ migration. 

\subsection{Long-Term Hybrid and Heterogeneous Operation}
All major PQ migration roadmaps, including those referenced by NIST and industry consortia, implicitly assume a multi-year and potentially even decade-long transition periods (for example, NIST provides the deadline of 2035 for the full migration of federal systems to PQC algorithms.) In such a scenario, best practice would be to assume that networks will operate in heterogeneous cryptographic states for an extended period of time, with PQ-capable components coexisting alongside classical, hybrid, and symmetric-only systems.

Indeed, if organizational risk posture demands ``quantum resistance now'' while remaining conservative about new primitives, then a best practice is to consider hybrid public-key mechanisms (e.g., classical + PQC) as a transitional tool. CSWP 39 explains the main operational rationale: continue using well-tested traditional public-key algorithms while PQC algorithms and implementations mature, recognizing that hybrids may also imply a second transition later. 

Rather than treating this heterogeneity as a temporary inconvenience, networks should be explicitly engineered to tolerate and embrace it. This includes clear downgrade resistance during hybrid operation, explicit policy boundaries around where PQ guarantees apply, and careful avoidance of ``weakest-link'' effects where legacy components potentially undermine PQ-secure segments. Systems that are only secure once everything is upgraded contradict both operational reality and NIST's emphasis on risk-managed, phased transition.

Furthermore, given the predominant risk of ``harvest now, decrypt later'' attacks, guidance from NIST and other roadmaps recommend shorter cryptoperiods even before large-scale quantum computers exist. At the network level, this means that keys must be short-lived, and that key rotation should itself be automated. In PQ-enabled systems, this means favoring ephemeral key establishment and frequent rekeying of long-lived tunnels or credentials. In symmetric-only architectures, where perfect forward secrecy is harder to achieve, this requires explicit mechanisms such as epoch-based rekeying, hash-chain evolution, or periodic KDS-assisted refresh. Architectures that cannot rotate keys at scale, due to reasons such as operational friction, manual processes, or static trust relationships, should be considered high-risk in PQ threat models, regardless of algorithm choice.

If repeated transitions are not possible (for example, due to embedded device lifetimes), then hybrids must be used with extra caution: CSWP 39 notes that some deployments may avoid a second transition by continuing to use a hybrid even when the classical component becomes disallowed, which can create governance and compliance problems. A best practice here is to encode ``exit conditions'' into policy and lifecycle management, not just into engineering intent.

\subsection{Design Explicitly for Post-Compromise Recovery}
Traditional security architectures often emphasize compromise prevention while treating recovery as an exceptional event. In the PQ era, this assumption is no longer tenable. NIST SP 800-208 explicitly acknowledges that some cryptographic material in use today may be compromised in the future, even if systems appear secure at present.

Therefore, best practice would be to design for post-compromise recovery and key healing right from the start. Threshold and MPC-backed KDS architectures, proactive resharing, and multi-path entropy injection are particularly valuable here, as they help to restrict the extent of the penetration and enable gradual recovery without global re-enrollment. Systems that lack recovery mechanisms implicitly assume perfect detection and instantaneous response, both of which are unrealistic in the realm of long-lived, stealthy PQ adversaries.

\subsection{Interoperability, Performance Testing, and Alignment with Risk Frameworks}
Given the increasing reliance on vendor ecosystems or multi-organization connectivity, it is important to treat interoperability testing as a central activity, not an afterthought. NCCoE describes interoperability testing as a dedicated workstream intended to identify compatibility issues, resolve them in controlled environments, and reduce duplicated effort across organizations. Thus, in operational terms it is important to run cross-vendor tests early, standardize test vectors and success criteria, and ensure that ``PQC-capable'' claims are validated at the protocol boundary (not just in isolated crypto libraries).

For organizations driven by risk governance (for example, audits, sector regulation, internal assurance), a best practice would be to express PQC migration activities in the language that security programs already use. NIST CSWP 48\footnote{https://nvlpubs.nist.gov/nistpubs/CSWP/NIST.CSWP.48.ipd.pdf} is explicitly a mapping document that connects the migration-to-PQC problem space to widely used risk frameworks (e.g., NIST CSF and SP 800-53 control families), which makes PQC plans easier to justify, track, and audit. Therefore, a best practice guidance would be to connect discovery, prioritization, testing, and rollout to existing risk and control structures so that migration work is not treated as an isolated crypto project.

%% file: ConcludingRemarks.tex
The transition to the PQ era poses challenges that extend well beyond the substitution of cryptographic primitives. As we have shown, quantum resilience in networks is an architectural and systems problem, influenced by cryptographic foundations, key-distribution mechanisms, trust assumptions, lifecycle management, deployment environments, and communication topology. Addressing PQ security as a protocol-local upgrade may risk missing structural vulnerabilities that arise from centralized trust, long-lived keys, limited recovery mechanisms, and heterogeneous operational constraints.

By presenting a unified taxonomy of PQ–resistant network architectures, this work clarifies the design space and highlights that no single approach dominates across all environments. Symmetric-only, PQ-PKI-based, hybrid, threshold/MPC-backed, and serverless multi-path architectures each offer distinct trade-offs in terms of security guarantees, scalability, deployability, and operational complexity. Making these trade-offs explicit is essential for informed decision-making, particularly in real-world systems that must operate under partial PQ readiness, regulatory constraints, and long migration timelines.

A central insight of our analysis is that cryptographic agility in networks is an emergent system property rather than just a software feature. The ability to adapt to evolving cryptographic assumptions depends critically on architectural choices that govern trust distribution, key lifecycle control, compromise containment, and recovery. Architectures that assume eventual compromise and support proactive rotation, threshold trust, and lifecycle-aware key management are inherently better positioned to withstand long-term quantum threats. We hope that this systematization provides both researchers and practitioners with a common framework for reasoning about quantum-resistant network design. By identifying fundamental gaps and promising research directions, in particular at the intersection of PQ cryptography, distributed systems, and network security, we aim to inform the development of robust, adaptable, and future-proof network infrastructures for the PQ era.

%% file: Appendix.tex
\begin{table*}[h]
\centering
\footnotesize
\renewcommand{\arraystretch}{1.0}
\caption{Acronyms Used in This Paper}
\label{tab:acronyms}
\begin{tabular}{|p{3.0cm}|p{10.0cm}|}
\hline
\textbf{Acronym} & \textbf{Meaning} \\
\hline
AES & Advanced Encryption Standard \\
\hline
AES-CTR & Advanced Encryption Standard with Counter Mode\\
\hline
AES-GCM & Advanced Encryption Standard with Galois/Counter Mode\\
\hline
AS & Autonomous System \\
\hline
BFT & Byzantine Fault Tolerance \\
\hline
CA & Certificate Authority \\
\hline
CRQC & Cryptographically Relevant Quantum Computer\\
\hline
DH & Diffie-Hellman\\
\hline
ECC & Elliptic Curve Cryptography \\
\hline
GDH & Group Diffie--Hellman \\
\hline
GKE & Group Key Establishment \\
\hline
HSM & Hardware Security Module \\
\hline
IKE & Internet Key Exchange \\
\hline
IPsec & Internet Protocol Security \\
\hline
KDC & Key Distribution Center \\
\hline
KDF & Key Derivation Function \\
\hline
KDS & Key Distribution Server \\
\hline
KEM & Key Encapsulation Mechanism \\
\hline
LKH & Logical Key Hierarchy \\
\hline
MAC & Message Authentication Code \\
\hline
MPC & Multi-Party Computation \\
\hline
MTU & Maximum Transmission Unit \\
\hline
ML-DSA & Module-Lattice-Based Digital Signature Algorithm \\
\hline
ML-KEM & Module-Lattice Encapsulation Mechanism \\
\hline
PFS & Perfect Forward Secrecy \\
\hline
PKC & Public-Key Cryptography \\
\hline
PKI & Public Key Infrastructure \\
\hline
PQ & Post-Quantum \\
\hline
PQC & Post-Quantum Cryptography \\
\hline
QRNG & Quantum Random Number Generator \\
\hline
QKD & Quantum Key Distribution \\
\hline
RPKI & Resource Public Key Infrastructure \\
\hline
TEE & Trusted Execution Environment \\
\hline
SSH & Secure Shell\\
\hline
TLS & Transport Layer Security \\
\hline
V2X & Vehicle-to-Everything Communication \\
\hline
\end{tabular}
\end{table*}

%% file: References.bib
@inproceedings{howe2021sok,
  title        = {SoK: How (Not) to Design and Implement Post-Quantum Cryptography},
  author       = {Howe, Jasmine and Prest, Thomas and Apon, Daniel},
  booktitle    = {Proceedings of CT-RSA 2021},
  series       = {Lecture Notes in Computer Science},
  year         = {2021},
  publisher    = {Springer},
  doi          = {10.1007/978-3-030-75539-3_11}
}

@article{alnahawi2023pqtlseprint,
  title        = {SoK: Post-Quantum TLS Handshake},
  author       = {Alnahawi, Nasser and Müller, Johannes and Oupick{\'y}, Jan and Wiesmaier, Alexander},
  journal      = {IACR Cryptology ePrint Archive},
  volume       = {2023},
  pages        = {1873},
  year         = {2023},
  url          = {https://eprint.iacr.org/2023/1873}
}

@article{alnahawi2024pqtlssurvey,
  title        = {A Comprehensive Survey on Post-Quantum TLS},
  author       = {Alnahawi, Nasser and Müller, Johannes and Oupick{\'y}, Jan and Wiesmaier, Alexander},
  journal      = {IACR Communications in Cryptology},
  year         = {2024},
  url          = {https://www.researchgate.net/publication/382087978_A_Comprehensive_Survey_on_Post-Quantum_TLS}
}

@article{naether2024agility,
  title        = {SoK: Towards a Common Understanding of Cryptographic Agility},
  author       = {N{\"a}ther, Christian and Herzinger, Daniel and Gazdag, Stefan-Lukas and Stegh{\"o}fer, Jan-Philipp and Daum, Simon and Loebenberger, Daniel},
  journal      = {arXiv preprint arXiv:2401.16443},
  year         = {2024},
  url          = {https://arxiv.org/abs/2401.16443}
}

@article{naether2024slr,
  title        = {Migrating Software Systems towards Post-Quantum Cryptography: A Systematic Literature Review},
  author       = {N{\"a}ther, Christian and Herzinger, Daniel and Gazdag, Stefan-Lukas and Stegh{\"o}fer, Jan-Philipp and Daum, Simon and Loebenberger, Daniel},
  journal      = {arXiv preprint arXiv:2404.12854},
  year         = {2024},
  url          = {https://arxiv.org/abs/2404.12854}
}

@article{kumibe2025hybrid,
  title        = {Post-Quantum Migration Strategies: A Hybrid Approach to Cryptographic Transition},
  author       = {Kumibe, Lamide and Oladele, Sunday},
  journal      = {ResearchGate Preprint},
  year         = {2025},
  url          = {https://www.researchgate.net/publication/392082187_Post-Quantum_Migration_Strategies_A_Hybrid_Approach_to_Cryptographic_Transition}
}

@article{Katsis,
  title        = {The Zero-trust Paradigm: Concepts, Architectures and Applications},
  author       = {Katsis, Charalampos and Bertino, Elisa},
  journal      = {Foundations and Trends in Privacy and Security},
  year         = {2025},
volume = {8},
number = {2},
  doi           = {10.1561/3300000046}
}

@article{node_capture_2014,
author = {Newell, Andrew and Yao, Hongyi and Ryker, Alex and Ho, Tracey and Nita-Rotaru, Cristina},
title = {Node-Capture Resilient Key Establishment in Sensor Networks: Design Space and New Protocols},
year = {2014},
issue_date = {January 2015},
publisher = {Association for Computing Machinery},
address = {New York, NY, USA},
volume = {47},
number = {2},
issn = {0360-0300},
url = {https://doi.org/10.1145/2636344},
doi = {10.1145/2636344},
journal = {ACM Comput. Surv.},
month = aug,
articleno = {24},
numpages = {34}}

@inproceedings{mpss_ndss_2021,
  title     = {{More than a Fair Share: Network Data Remanence Attacks against Secret Sharing-based Schemes}},
  author    = {Leila Rashidi and Daniel Kostecki and Alexander James and Anthony Peterson and Majid Ghaderi and Samuel Jero and Cristina Nita-Rotaru and Hamed Okhravi and Reihaneh Safavi-Naini},
  booktitle = {Proceedings of the 28th Annual Network and Distributed System Security Symposium (NDSS 2021)},
  year      = {2021},
  publisher = {The Internet Society},
  doi       = {10.14722/ndss.2021.23062},
  url       = {https://www.ndss-symposium.org/wp-content/uploads/ndss2021_2C-2_23062_paper.pdf}
}

@inproceedings{butler2005kerberos-crossrealm,
  author    = {F. Butler and I. Cervesato and A. D. Jaggard and A. Scedrov},
  title     = {Specifying Kerberos 5 Cross-Realm Authentication},
  booktitle = {Workshop on Issues in the Theory of Security (WITS)},
  year      = {2005},
  pages     = {12--26},
  organization = {ACM}
}

@techreport{nist2024pqc-standards,
  author      = {{National Institute of Standards and Technology}},
  title       = {FIPS 203, 204, 205: Module-Lattice-Based Key-Encapsulation Mechanism, Digital Signature Algorithm, and Stateless Hash-Based Digital Signature Algorithm},
  institution = {NIST},
  year        = {2024},
  month       = {August}
}

@article{techrxiv2025pqc-zt,
  author  = {M. Kumar and others},
  title   = {Post-Quantum Cryptographic Integration Framework for Zero Trust Enterprise Cloud Environments},
  journal = {TechRxiv Preprint},
  year    = {2025},
  month   = {December},
  doi     = {10.36227/techrxiv.1369271}
}

@article{dervisevic2024qkd-survey,
  author  = {E. Dervisevic and A. Tankovic and E. Fazel and R. Kompella and P. Fazio and M. Voznak and M. Mehic},
  title   = {Quantum Key Distribution Networks -- Key Management: A Survey},
  journal = {ACM Computing Surveys},
  year    = {2025},
  month   = {May}
}

@techreport{ietf2025composite-guidance,
  author      = {T. Reddy and D. Migault and H. Tschofenig},
  title       = {Guidance for Migration to Composite, Dual, or PQC Authentication},
  institution = {IETF},
  type        = {Internet-Draft},
  number      = {draft-reddy-pquip-pqc-signature-migration},
  year        = {2025},
  month       = {October}
}

@techreport{ietf2025kem-ikev2,
  author      = {M. Wang and others},
  title       = {KEM-based Authentication for IKEv2 with Post-quantum Security},
  institution = {IETF},
  type        = {Internet-Draft},
  number      = {draft-wang-ipsecme-kem-auth-ikev2-02},
  year        = {2025},
  month       = {October}
}

@misc{redhat2025pqc-kerberos,
  author       = {{Red Hat Research}},
  title        = {The Post-Quantum Cryptography Transition: PQC Linux Authentication Pilot},
  howpublished = {Red Hat Research Blog},
  year         = {2025},
  month        = {October},
  note         = {MIT Kerberos quantum-safe adaptation research}
}

@article{adan2025quantum-auth,
  author  = {A. Adan and others},
  title   = {Quantum-resistant Authentication: Securing Identity and Data Against Future Threats},
  journal = {AIMS Mathematics},
  year    = {2025},
  volume  = {10},
  number  = {8},
  pages   = {779--810},
  month   = {July}
}

@article{arxiv2025applied-pqc-pki,
  author  = {P. Kampanakis and others},
  title   = {Applied Post Quantum Cryptography: A Practical Approach to Hybrid Certificate Management},
  journal = {arXiv preprint arXiv:2505.04333},
  year    = {2025},
  month   = {May}
}

@inproceedings{thinkmind2024composite-certs,
  author    = {M. Varga and others},
  title     = {Theoretical and Practical Aspects in Identifying Gaps During Post-Quantum Migration},
  booktitle = {Proceedings of SECURWARE 2024},
  year      = {2024},
  pages     = {20--30},
  publisher = {ThinkMind}
}

@techreport{nist2025crypto-agility,
  author      = {{NIST NCCoE}},
  title       = {Migration to Post-Quantum Cryptography: Preparation for Considering the Implementation and Adoption of Quantum Safe Cryptography},
  institution = {NIST National Cybersecurity Center of Excellence},
  year        = {2025},
  note        = {SP 1800-38 Draft}
}

@inproceedings{wallner1999rfc2627,
  author    = {D. M. Wallner and E. J. Harder and R. C. Agee},
  title     = {Key Management for Multicast: Issues and Architectures},
  booktitle = {RFC 2627},
  year      = {1999},
  publisher = {IETF}
}

@inproceedings{RFC6806,
  author    = {S. Hartman and K. Raeburn and L. Zhu},
  title     = {Kerberos Principal Name Canonicalization and Cross-Realm Referrals},
  booktitle = {RFC 6806},
  year      = {1999},
  publisher = {IETF}
}

@inproceedings{Kampanakis,
  author    = {P. Kampanakis},
  title     = {Post-quantum Hybrid Key Exchange with ML-KEM in the Internet Key Exchange Protocol Version 2 (IKEv2)},
  booktitle = {Proposed Standard},
  year      = {2025},
  publisher = {IETF}
}

@inproceedings{RFC4556,
  author    = {L. Zhu and B. Tung},
  title     = {Public Key Cryptography for
              Initial Authentication in Kerberos (PKINIT)},
  booktitle = {RFC 4556},
  year      = {2006},
  publisher = {IETF}
}

@article{mcgrew2003oft,
  author  = {D. A. McGrew and A. T. Sherman},
  title   = {Key Establishment in Large Dynamic Groups Using One-Way Function Trees},
  journal = {IEEE Transactions on Software Engineering},
  year    = {2003},
  volume  = {29},
  number  = {5},
  pages   = {444--458}
}

@article{Souvatzidaki,
  author  = {K. Souvatzidaki and K. Limniotis},
  title   = {Post-Quantum Key Exchange in TLS 1.3: Further Analysis on Performance of New Cryptographic Standards},
  journal = {Cryptography},
  year    = {2025},
  volume  = {9},
  number  = {4},
  pages   = {2-24}
}

@inproceedings{mittra1997iolus,
  author    = {S. Mittra},
  title     = {Iolus: A Framework for Scalable Secure Multicasting},
  booktitle = {Proceedings of ACM SIGCOMM},
  year      = {1997},
  pages     = {277--288}
}

@article{wong2000keygraphs,
  author  = {C. K. Wong and M. Gouda and S. S. Lam},
  title   = {Secure Group Communications Using Key Graphs},
  journal = {IEEE/ACM Transactions on Networking},
  year    = {2000},
  volume  = {8},
  number  = {1},
  pages   = {16--30}
}

@techreport{harney1997gkmp,
  author      = {H. Harney and C. Muckenhirn},
  title       = {Group Key Management Protocol (GKMP) Architecture},
  institution = {IETF},
  type        = {RFC},
  number      = {2094},
  year        = {1997}
}

@techreport{ietf-tls-kdh,
  author      = {R. {Van Rein}},
  title       = {Quantum Relief with TLS and Kerberos},
  institution = {IETF},
  type        = {Internet-Draft},
  number      = {draft-vanrein-tls-kdh-08},
  year        = {2022}
}

@article{aldarwbi2020keyshield,
  author  = {M. Y. Al-Darwbi and A. A. Ghorbani and A. H. Lashkari},
  title   = {KeyShield: A Scalable and Quantum-Safe Key Management Scheme},
  journal = {IEEE Open Journal of the Communications Society},
  year    = {2020},
  volume  = {1},
  pages   = {1--14}
}

@article{ghosh2024scada-multiphase,
  author  = {S. Ghosh and M. Zaman and R. Joshi and S. Sampalli},
  title   = {Multi-Phase Quantum Resistant Framework for Secure Communication in SCADA Systems},
  journal = {IEEE Transactions on Dependable and Secure Computing},
  year    = {2024},
  volume  = {21},
  number  = {3}
}

@article{liu2024degkm,
  author  = {G. Liu and H. Li and N. Wang and T. Xiang and Y. Liu},
  title   = {DeGKM: Decentralized Group Key Management for Content Push in Integrated Networks},
  journal = {IEEE Transactions on Quantum Engineering},
  year    = {2024},
  volume  = {5}
}

@article{abdmeziem2025scada-taxonomy,
  author  = {A. Abdmeziem and A. Ahmed Nacer and N. Deroues},
  title   = {A Taxonomy of Key Management Schemes of SCADA Systems},
  journal = {IEEE Access},
  year    = {2025},
  volume  = {13},
  pages   = {1--22}
}

@article{dinker2025wsn,
  author  = {A. G. Dinker and V. Sharma},
  title   = {Exploring Cryptographic Key Management Schemes for Enhanced Security in Wireless Sensor Networks},
  journal = {Intelligent Automation and Soft Computing},
  year    = {2025},
  volume  = {37},
  number  = {3}
}

@techreport{fsisac2025pqc,
  author      = {{FS-ISAC PQC Working Group}},
  title       = {Post-Quantum Cryptography Working Group: Current State of Quantum Readiness},
  institution = {Financial Services Information Sharing and Analysis Center},
  year        = {2025}
}

@article{smith2024_5g6g,
  author  = {J. Smith and others},
  title   = {A Systematic Survey on 5G and 6G Security Considerations},
  journal = {Electrical Engineering Faculty Publications, University of Nebraska},
  year    = {2024}
}

@article{bella2025macsec,
  author  = {M. Bella and P. Marchetta},
  title   = {Robust Multicast Origin Authentication in MACsec and CANsec for Automotive Networks},
  journal = {arXiv preprint arXiv:2502.20555},
  year    = {2025}
}

@article{zhang2025iot-survey,
  author  = {K. Zhang and others},
  title   = {A Comprehensive Survey on Cybersecurity Threats and Defense Mechanisms in IoT},
  journal = {arXiv preprint arXiv:2601.00556},
  year    = {2025}
}

@article{kumar2024pq-cloud,
  author  = {P. Kumar and R. S. Rao},
  title   = {Post-Quantum Cryptography for Secure Data Transmission in Cloud Environments},
  journal = {TIJER International Journal},
  year    = {2024},
  volume  = {11},
  number  = {12}
}

@phdthesis{nita2001phd,
  author = {Cristina Nita-Rotaru},
  title  = {High-Performance Secure Group Communication},
  school = {Johns Hopkins University},
  year   = {2001},
  note   = {Advisor: Yair Amir}
}

@article{amir2004robust-gka,
  author  = {Yair Amir and Yongdae Kim and Cristina Nita-Rotaru and John L. Schultz and Jonathan Stanton and Gene Tsudik},
  title   = {Secure Group Communication Using Robust Contributory Key Agreement},
  journal = {IEEE Transactions on Parallel and Distributed Systems},
  year    = {2004},
  volume  = {15},
  number  = {5},
  pages   = {468--480},
  doi     = {10.1109/TPDS.2004.1278099}
}

@inproceedings{nita2002cliques-dsn,
  author    = {Cristina Nita-Rotaru and Yair Amir and Jonathan Stanton and Gene Tsudik},
  title     = {Secure Group Communication in Asynchronous Networks with Failures: Integration and Experiments},
  booktitle = {Proceedings of the 20th IEEE International Conference on Distributed Computing Systems (ICDCS)},
  year      = {2000},
  pages     = {330--343},
  doi       = {10.1109/ICDCS.2000.840943}
}

@article{kim2004tgdh-acm,
  author  = {Yongdae Kim and Adrian Perrig and Gene Tsudik},
  title   = {Tree-Based Group Key Agreement},
  journal = {ACM Transactions on Information and System Security (TISSEC)},
  year    = {2004},
  volume  = {7},
  number  = {1},
  pages   = {60--96},
  doi     = {10.1145/984334.984337}
}

@inproceedings{lee2002tgdh-icnp,
  author    = {Patrick P. C. Lee and John C. S. Lui and David K. Y. Yau},
  title     = {Distributed Collaborative Key Agreement Protocols for Dynamic Peer Groups},
  booktitle = {Proceedings of IEEE ICNP 2002},
  year      = {2002},
  pages     = {92--101},
  doi       = {10.1109/ICNP.2002.1181388}
}

@inproceedings{buttyan2012invitation-tgdh,
  author    = {Levente Butty\'{a}n and Tam\'{a}s Holczer and P\'{e}ter Szil\'{a}gyi},
  title     = {Invitation-Oriented TGDH: Key Management for Dynamic Groups in the Cloud},
  booktitle = {Proceedings of CloudSec 2012},
  year      = {2012},
  publisher = {Springer},
  pages     = {1--15}
}

@article{pablos2020kyber-gake,
  author  = {\'{A}lvaro Pablos Cantos and Mar\'{\i}a Isabel Gonz\'{a}lez Vasco and Misael Enrique Mart\'{\i}nez P\'{e}rez and Rainer Steinwandt},
  title   = {Compiled Constructions towards Post-Quantum Group Key Exchange: A Design from Kyber},
  journal = {Mathematics},
  year    = {2020},
  volume  = {8},
  number  = {10},
  pages   = {1853},
  doi     = {10.3390/math8101853},
  publisher = {MDPI}
}

@article{pablos2022iet-kyber,
  author  = {\'{A}lvaro Pablos Cantos and Mar\'{\i}a Isabel Gonz\'{a}lez Vasco and Rainer Steinwandt},
  title   = {Secure Post-Quantum Group Key Exchange from Kyber},
  journal = {IET Information Security},
  year    = {2022},
  volume  = {16},
  number  = {5},
  pages   = {561--571},
  doi     = {10.1049/cmu2.12561}
}

@article{zhang2023lattice-gka-tworounds,
  author  = {Xiuhua Zhang and Yiliang Han and Xu An Wang and Xiaoyuan Yang},
  title   = {A Two Rounds Dynamic Authenticated Group Key Agreement Protocol Based on LWE},
  journal = {Journal of Information Security and Applications},
  year    = {2023},
  volume  = {73},
  pages   = {103424},
  doi     = {10.1016/j.jisa.2022.103424}
}

@inproceedings{li2024lattice-dgka,
  author    = {Yang Li and Jianghong Wei and Guomin Yang and Willy Susilo},
  title     = {A Lattice-Based Dynamic Group Authenticated Key Exchange Protocol},
  booktitle = {Proceedings of IEEE TrustCom 2024},
  year      = {2024},
  pages     = {1--8},
  doi       = {10.1109/TrustCom63042.2024.00015}
}

@article{vasco2025scalable-qkd-hybrid,
  author  = {Mar\'{\i}a Isabel Gonz\'{a}lez Vasco and Rainer Steinwandt},
  title   = {Scalable Authenticated Group Key Establishment in Quantum and Post-Quantum Networks},
  journal = {Informatica},
  year    = {2025},
  volume  = {36},
  number  = {2},
  pages   = {299--320},
  doi     = {10.15388/25-INFOR549}
}

@article{zhang2025v2x-lightweight,
  author  = {Yifan Zhang and Jinhua Zheng and Xuan Wang and Jian Chen and Shaohua Wan},
  title   = {A Lightweight Key Agreement Protocol for V2X Communications Based on Hybrid Post-Quantum Cryptography},
  journal = {Sensors},
  year    = {2025},
  volume  = {25},
  number  = {22},
  pages   = {7239},
  doi     = {10.3390/s25227239}
}

@article{yadav2024blockchain-vanet,
  author  = {Abhishek Yadav and Pradeep Kumar Singh and Narendra Singh and Gautam Srivastava},
  title   = {Module Lattice Based Post Quantum Secure Blockchain Empowered Authentication Framework for Autonomous Vehicle Communication System},
  journal = {Computers and Electrical Engineering},
  year    = {2024},
  volume  = {119},
  pages   = {109442},
  doi     = {10.1016/j.compeleceng.2024.109442}
}

@article{wu2025pq-vanet-mac,
  author  = {Hao Wu and Qinglei Kong and Zheng Gong and Muhammad Waqas},
  title   = {Post-Quantum Weighted Anonymous Authentication for Hybrid VANET MAC Protocol},
  journal = {IEEE Transactions on Vehicular Technology},
  year    = {2025},
  volume  = {74},
  number  = {7},
  pages   = {9143--9157},
  doi     = {10.1109/TVT.2025.3385647}
}

@article{demir2024uav-pqc-survey,
  author  = {Esra Demir and Muhammad Waqas and Zheng Gong and Gautam Srivastava},
  title   = {Future-Proofing Security for UAVs with Post-Quantum Cryptography: A Review},
  journal = {IEEE Access},
  year    = {2024},
  volume  = {12},
  pages   = {165321--165349},
  doi     = {10.1109/ACCESS.2024.3495832}
}

@article{YavuzAlagozAnarim2010,
  author  = {Yavuz, Attila A. and Alag{\"o}z, Fatih and Anarim, Emin},
  title   = {A New Multi-Tier Adaptive Military MANET Security Protocol Using Hybrid Cryptography and Signcryption},
  journal = {Turkish Journal of Electrical Engineering and Computer Sciences},
  volume  = {18},
  number  = {1},
  pages   = {1--22},
  year    = {2010},
  doi     = {10.3906/elk-0904-6}
}

@article{alhaj2025uav-mavlink-pqc,
  author  = {Hussein Al-Haj and Zaher Al Bahou and Ammar Kadi and Ahmad Musa},
  title   = {Post-Quantum Cryptography for Military UAV Communication Systems: Integration with MAVLink Protocol},
  journal = {Authorea Preprints},
  year    = {2025},
  month   = {November},
  doi     = {10.22541/au.176281850.08964861}
}

@article{scidir2024uav-swarm-survey,
  author  = {Menglin Chu and Hao Wang and Jun Zhao and Zhangjie Fu},
  title   = {Intelligent UAV Swarm Key Agreement Survey: Systematic Taxonomy and Analysis},
  journal = {Internet of Things},
  year    = {2025},
  volume  = {30},
  pages   = {101449},
  doi     = {10.1016/j.iot.2025.101449}
}

@article{nature2025heterogeneous-qkn,
  author  = {Yaxing Chen and Tianqi Dou and Qiang Zhou and Qiong Li},
  title   = {Fully Heterogeneous Prepare-and-Measure Quantum Network for the Quantum Internet},
  journal = {Nature Communications},
  year    = {2025},
  volume  = {16},
  pages   = {966},
  doi     = {10.1038/s41467-025-66333-3}
}

@misc{nict2025qkd-multiplex,
  author       = {{NICT Japan}},
  title        = {World's First Integrated System for Quantum Key Distribution Network on Backbone Optical Network},
  howpublished = {NICT Press Release},
  year         = {2025},
  month        = {September},
  note         = {Multiple QKD protocol coexistence demonstration}
}

@inproceedings{castryck2018csidh,
  author    = {Wouter Castryck and Tanja Lange and Chloe Martindale and Lorenz Panny and Joost Renes},
  title     = {CSIDH: An Efficient Post-Quantum Commutative Group Action},
  booktitle = {Proceedings of ASIACRYPT 2018},
  year      = {2018},
  pages     = {395--427},
  publisher = {Springer},
  doi       = {10.1007/978-3-030-03332-3_15}
}

@misc{kuleuven2018csidh,
  author       = {Wouter Castryck and Tanja Lange},
  title        = {CSIDH: Post-Quantum Key Exchange Using Isogeny-Based Group Actions},
  howpublished = {KU Leuven COSIC Research Blog},
  year         = {2018},
  month        = {December}
}

@techreport{ietf2025pqc-engineers,
  author      = {Amir Sehovic and Sofía Celi and Thom Wiggers},
  title       = {Post-Quantum Cryptography for Engineers},
  institution = {IETF},
  type        = {Internet-Draft},
  number      = {draft-ietf-pquip-pqc-engineers-13},
  year        = {2025},
  month       = {June}
}

@inproceedings{burmester1994secure,
  title={A secure and efficient conference key distribution system},
  author={Burmester, Mike and Desmedt, Yvo},
  booktitle={Advances in Cryptology---EUROCRYPT '94},
  pages={275--286},
  year={1994},
  organization={Springer}
}

@inproceedings{katz2003scalable,
  title={Scalable protocols for authenticated group key exchange},
  author={Katz, Jonathan and Yung, Moti},
  booktitle={Advances in Cryptology---CRYPTO 2003},
  pages={110--125},
  year={2003},
  organization={Springer}
}

@inproceedings{campagna2013kerberos,
  author    = {Campagna, Matt and Hardjono, Thomas and Pintsov, Leon and Romansky, Boris and Yu, Tom},
  title     = {Kerberos Revisited: Quantum-Safe Authentication},
  booktitle = {ETSI Quantum-Safe-Crypto Workshop},
  year      = {2013},
  month     = sep,
  address   = {Sophia Antipolis, France},
  url       = {https://docbox.etsi.org/Workshop/2013/201309_CRYPTO/S03_INDUSTRY_SESSION/PITNEYBOWES_PINTSOV.pdf},
  note      = {Workshop presentation (slides)}
}

@article{baseri2024navigating,
title = {Navigating quantum security risks in networked environments: A comprehensive study of quantum-safe network protocols},
journal = {Computers and Security},
volume = {142},
pages = {103883},
year = {2024},
issn = {0167-4048},
doi = {https://doi.org/10.1016/j.cose.2024.103883},
url = {https://www.sciencedirect.com/science/article/pii/S0167404824001846},
author = {Yaser Baseri and Vikas Chouhan and Abdelhakim Hafid}}

@inproceedings{sherman2014needham,
  author    = {Sherman, Alan T.},
  title     = {Needham--Schroeder, Kerberos, and Quantum Computing},
  booktitle = {Course handout / lecture notes},
  year      = {2018},
  url       = {https://courses.cs.umbc.edu/undergraduate/426/spring18/schedule/network/kerberos_quantum_handout.pdf}}

@techreport{nist2024fips203,
  author      = {{National Institute of Standards and Technology}},
  title       = {Module-Lattice-Based Key-Encapsulation Mechanism Standard},
  institution = {National Institute of Standards and Technology},
  type        = {Federal Information Processing Standards Publication},
  number      = {203},
  year        = {2024},
  month       = aug,
  note        = {Published Aug. 2024}
}

@techreport{nist2024fips204,
  author      = {{National Institute of Standards and Technology}},
  title       = {Module-Lattice-Based Digital Signature Standard},
  institution = {National Institute of Standards and Technology},
  type        = {Federal Information Processing Standards Publication},
  number      = {204},
  year        = {2024},
  month       = aug,
  url         = {https://nvlpubs.nist.gov/nistpubs/fips/nist.fips.204.pdf},
  note        = {Published Aug. 2024}
}

@article{redhat2025pqc,
  author  = {Hinds, Luke},
  title   = {The Post-Quantum Cryptography Transition: Researching a Quantum-Safe Future},
  journal = {Red Hat Research},
  year    = {2025},
  url     = {https://research.redhat.com/blog/article/the-post-quantum-cryptography-transition-researching-a-quantum-safe-future/}}

@misc{rfc9794,
  author       = {Florence, F. D. D. and Ounsworth, Mike and Mister, Scott},
  title        = {Terminology for Post-Quantum Traditional Hybrid Schemes},
  howpublished = {IETF RFC 9794},
  year         = {2025},
  month        = jun,
  url          = {https://datatracker.ietf.org/doc/html/rfc9794}
}

@techreport{mskkdcp,
  author      = {{Microsoft Corporation}},
  title       = {[MS-KKDCP]: Kerberos Key Distribution Center (KDC) Proxy Protocol},
  institution = {Microsoft Open Specifications},
  year        = {2024},
  url         = {https://learn.microsoft.com/en-us/openspecs/windows_protocols/ms-kkdcp/},
  note        = {Open specification for tunneling Kerberos to a proxy over HTTP(S)}
}

@INPROCEEDINGS{zhang2025lattice,
  author={Zhang, Rui and Chen, Liquan and Liu, Suhui and Fang, Huiyu and Lu, Yan},
  booktitle={2024 10th International Conference on Computer and Communications (ICCC)}, 
  title={A Lattice-Based Dynamic Group Authenticated Key Exchange Protocol}, 
  year={2024},
  volume={},
  number={},
  pages={387-393},
  doi={10.1109/ICCC62609.2024.10942140}}

@article{cao2025lpbt,
  author  = {Cao, Chenchen and Xu, Chunxiang and Jiang, Changsong and Zhang, Zhao and Dong, Xinfeng},
  title   = {LPbT-SSO: Password-Based Threshold Single-Sign-On Authentication From LWE},
  journal = {IEEE Transactions on Dependable and Secure Computing},
  year    = {2025},
  url     = {https://www.computer.org/csdl/journal/tq/2025/01},
  note    = {Metadata (volume/issue/DOI/pages) should be pulled from the IEEE record for precision}
}

@inproceedings{castro1999pbft,
  author    = {Miguel Castro and Barbara Liskov},
  title     = {Practical Byzantine Fault Tolerance},
  booktitle = {Proceedings of the Third Symposium on Operating Systems Design and Implementation (OSDI '99)},
  year      = {1999},
  publisher = {USENIX Association},
  url       = {https://www.usenix.org/conference/osdi-99/practical-byzantine-fault-tolerance}
}

@inproceedings{castro2000proactive,
  author    = {Miguel Castro and Barbara Liskov},
  title     = {Proactive Recovery in a Byzantine-Fault-Tolerant System},
  booktitle = {Proceedings of the Fourth Symposium on Operating Systems Design and Implementation (OSDI 2000)},
  year      = {2000},
  publisher = {USENIX Association},
  url       = {https://www.usenix.org/events/osdi2000/castro/castro.pdf}
}

@INPROCEEDINGS{10063497,
  author={Yavuz, Attila A. and Nouma, Saif E. and Hoang, Thang and Earl, Duncan and Packard, Scott},
  booktitle={2022 IEEE 4th International Conference on Trust, Privacy and Security in Intelligent Systems, and Applications (TPS-ISA)}, 
  title={Distributed Cyber-infrastructures and Artificial Intelligence in Hybrid Post-Quantum Era}, 
  year={2022},
  pages={29-38},
  doi={10.1109/TPS-ISA56441.2022.00014}}

@article{shamir1979,
  author  = {Adi Shamir},
  title   = {How to Share a Secret},
  journal = {Communications of the ACM},
  volume  = {22},
  number  = {11},
  pages   = {612--613},
  year    = {1979},
  publisher = {ACM},
  doi     = {10.1145/359168.359176},
  url     = {https://dl.acm.org/doi/abs/10.1145/359168.359176}
}

@inproceedings{chuang1997pkda,
  author    = {John Chung-I Chuang and Marvin A. Sirbu},
  title     = {Distributed Authentication in Kerberos Using Public Key Cryptography},
  booktitle = {Proceedings of the Network and Distributed System Security Symposium (NDSS '97)},
  year      = {1997},
  url       = {https://people.ischool.berkeley.edu/~chuang/pubs/pkda.pdf},
  note      = {Also referred to as PKDA (public-key based Kerberos for distributed authentication)}
}

@inproceedings{araki2016ht3pc,
  author    = {Toshinori Araki and Jun Furukawa and Yehuda Lindell and Ariel Nof and Kazuma Ohara},
  title     = {High-Throughput Semi-Honest Secure Three-Party Computation with an Honest Majority},
  booktitle = {Proceedings of the 2016 ACM SIGSAC Conference on Computer and Communications Security (CCS '16)},
  year      = {2016},
  publisher = {ACM},
  doi       = {10.1145/2976749.2978331},
  url       = {https://dl.acm.org/doi/10.1145/2976749.2978331},
  note      = {Includes a Kerberos authentication / ticket-generation demonstration workload}
}

@article{10.1145/3772274,
author = {Sedghighadikolaei, Kiarash and Yavuz, Attila Altay},
title = {A Survey of Threshold Signatures: NIST Standards, Post-Quantum Cryptography, Exotic Techniques, and Real-World Applications},
year = {2025},
issue_date = {April 2026},
publisher = {Association for Computing Machinery},
address = {New York, NY, USA},
volume = {58},
number = {6},
issn = {0360-0300},
url = {https://doi.org/10.1145/3772274},
doi = {10.1145/3772274},
journal = {ACM Comput. Surv.},
month = dec,
articleno = {143},
numpages = {39}}

@inproceedings{agrawal2018dise,
  author    = {Shashank Agrawal and Payman Mohassel and Pratyay Mukherjee and Peter Rindal},
  title     = {DiSE: Distributed Symmetric-key Encryption},
  booktitle = {Proceedings of the 2018 ACM SIGSAC Conference on Computer and Communications Security (CCS '18)},
  year      = {2018},
  pages     = {1993--2010},
  publisher = {ACM},
  url       = {https://dblp.org/rec/conf/ccs/AgrawalMMR18}
}

@inproceedings{damgard2012spdz,
  author    = {Ivan Damg{\aa}rd and Valerio Pastro and Nigel P. Smart and Sarah Zakarias},
  title     = {Multiparty Computation from Somewhat Homomorphic Encryption},
  booktitle = {Advances in Cryptology -- CRYPTO 2012},
  year      = {2012},
  url       = {https://research-information.bris.ac.uk/en/publications/multiparty-computation-from-somewhat-homomorphic-encryption/}
}

@misc{cramer2018spdz2k,
  author = {Ronald Cramer and Ivan Damg{\aa}rd and Daniel Escudero and Peter Scholl and Chaoping Xing},
  title  = {{SPD $Z2^k$: Efficient MPC mod $2^k$ for Dishonest Majority}},
  year   = {2018},
  howpublished = {Cryptology ePrint Archive / CRYPTO 2018 preprint},
  url    = {https://pdfs.semanticscholar.org/6d73/fd19baf0b2e04b06749e5f8e42535e49270e.pdf}
}

@misc{keller2016mascot,
  author = {Marcel Keller and Emmanuela Orsini and Peter Scholl},
  title  = {MASCOT: Faster Malicious Arithmetic Secure Computation with Oblivious Transfer},
  year   = {2016},
  howpublished = {IACR Cryptology ePrint Archive},
  url    = {https://dblp.org/rec/journals/iacr/KellerOS16}
}

@misc{Goertzen,
author = {J. Goertzen and D. Stebila},
title= {Post-Quantum Signatures in DNSSEC via Request-Based Fragmentation},
year = {2022},
howpublished = {ArXiV},
  url    = { 	
https://doi.org/10.48550/arXiv.2211.14196}
}

@misc{pq_fingerprinting_2025,
      title={Classifying Implementations of Cryptographic Primitives and Protocols that Use Post-Quantum Algorithms}, 
      author={Tushin Mallick and Cristina Nita-Rotaru and Ashish Kundu and Ramana Kompella},
      year={2025},
      eprint={2503.17830},
      archivePrefix={arXiv},
      primaryClass={cs.CR},
      url={https://arxiv.org/abs/2503.17830}, 
}

@article{coca_2002,
author = {Zhou, Lidong and Schneider, Fred B. and Van Renesse, Robbert},
title = {COCA: A secure distributed online certification authority},
year = {2002},
issue_date = {November 2002},
publisher = {Association for Computing Machinery},
address = {New York, NY, USA},
volume = {20},
number = {4},
issn = {0734-2071},
url = {https://doi.org/10.1145/571637.571638},
doi = {10.1145/571637.571638},
journal = {ACM Trans. Comput. Syst.},
month = nov,
pages = {329–368},
numpages = {40}
}

@inproceedings{LouFang2001Multipath,
  author    = {Wenjing Lou and Yuguang Fang},
  title     = {A multipath routing approach for secure data delivery},
  booktitle = {Proceedings of MILCOM 2001 Communications for Network-Centric Operations: Creating the Information Force},
  year      = {2001},
  volume    = {2},
  pages     = {1467--1473},
  month     = oct,
  note      = {Cat. No. 01CH37277}
}

@article{AhmadiSafaviNaini2014Multipath,
  author  = {H. Ahmadi and R. Safavi-Naini},
  title   = {Multipath private communication: An information theoretic approach},
  journal = {arXiv preprint arXiv:1401.3659},
  year    = {2014}
}

@patent{DolevTzurDavid2017SecureInterconnection,
  author      = {S. Dolev and S. Tzur-David},
  title       = {A method for establishing a secure private interconnection over a multipath network},
  number      = {EP3146668A1},
  year        = {2017},
  month       = mar,
  assignee    = {European Patent Office}
}

@inproceedings{SafaviNainiPoostindouzLisy2017PathHopping,
  author    = {R. Safavi-Naini and A. Poostindouz and V. Lisy},
  title     = {Path hopping: An MTD strategy for quantum-safe communication},
  booktitle = {Proceedings of the ACM Workshop on Moving Target Defense},
  year      = {2017},
  pages     = {111--114}
}

@misc{mallick_quantum_blockchains_sok_2025,
      title={Quantum Disruption: An {SOK} of How Post-Quantum Attackers Reshape Blockchain Security and Performance}, 
      author={Tushin Mallick and Maya Zeldin and Murat Cenk and Cristina Nita-Rotaru},
      year={2025},
      eprint={2512.13333},
      archivePrefix={arXiv},
      primaryClass={cs.CR},
      url={https://arxiv.org/abs/2512.13333}, 
}

@inproceedings{basin_iMessage_usenix2025,
author = {Linker, Felix and Sasse, Ralf and Basin, David},
title = {A formal analysis of apple's iMessage PQ3 protocol},
year = {2025},
isbn = {978-1-939133-52-6},
publisher = {USENIX Association},
address = {USA},
booktitle = {Proceedings of the 34th USENIX Conference on Security Symposium},
articleno = {258},
numpages = {20},
location = {Seattle, WA, USA},
series = {SEC '25}
}

@inproceedings{pqxdh_usenix2024,
author = {Bhargavan, Karthikeyan and Jacomme, Charlie and Kiefer, Franziskus and Schmidt, Rolfe},
title = {Formal verification of the PQXDH post-quantum key agreement protocol for end-to-end secure messaging},
year = {2024},
isbn = {978-1-939133-44-1},
publisher = {USENIX Association},
address = {USA},
booktitle = {Proceedings of the 33rd USENIX Conference on Security Symposium},
articleno = {27},
numpages = {18},
location = {Philadelphia, PA, USA},
series = {SEC '24}
}

@misc{signal_spqr_2025,
  author       = {{Signal Messenger}},
  title        = {Signal Protocol and Post-Quantum Ratchets (SPQR)},
  year         = {2025},
  month        = October,
  howpublished = {\url{https://signal.org/blog/spqr/}},
  note         = {Blog post},
}
